# Towards generalized noise-level dependent crystallographic symmetry classifications of more or less periodic crystal patterns

Peter Moeck

Nano-Crystallography Group, Department of Physics, Portland State University, Portland, OR 97207-0751, USA,
pmoeck@pdx.edu

Geometric Akaike Information Criteria (G-AICs) for generalized noise-level dependent crystallographic symmetry classifications of two-dimensional (2D) images that are more or less periodic in either two or one dimensions as well as Akaike weights for multi-model inferences and predictions are reviewed. Such novel classifications do not refer to a single crystallographic symmetry class exclusively in a qualitative and definitive way. Instead, they are quantitative, spread over a range of crystallographic symmetry classes, and provide opportunities for inferences from all classes (within the range) simultaneously. The novel classifications are based on information theory and depend only on information that has been extracted from the images themselves by means of maximal likelihood approaches so that these classifications are objective. This is in stark contrast to the common practice whereby arbitrarily set thresholds are employed to force crystallographic symmetry classifications into apparently definitive/exclusive states, while the geometric feature extraction results on which they depend are never definitive in the presence of generalized noise, i.e. in all real world applications. Thus, there is unnecessary subjectivity in the currently practiced ways of making crystallographic symmetry classifications, which can be overcome by the approach outlined in this review.

**Keywords**: Crystallographic symmetry classifications, Geometric Akaike Information Criteria, Akaike weights, multi-model inference, information theory

## 1. Introduction and Background

While there is a large variety of extraction algorithms for geometric features such as point and translation symmetries from gray level patterns that are more or less periodic in two (2D) and one (1D) dimensions [1,2], related comments by Kenichi Kanatani [3] on symmetry as a continuous and hierarchic feature have been largely ignored for the last two decades by the computational symmetry and applied crystallography communities alike. The notable exceptions in this respect are the work by Yanxi Liu and coworkers [4,5] on 1D periodic time series in the form of subsequently recorded 2D images which were done more than a decade ago and much more recent work by the author of this review on objective 2D Bravais lattice type assignments to noisy images [6].

While the applied crystallography community typically speaks of "crystal patterns" when it refers to atomically resolved images [7], more or less 2D and 1D periodic patterns where the individual pixels possess digitized intensity values (i.e. gray-levels rather than colors) are commonly referred to as "near regular textures" within the computational symmetry community [1]. With the title of this review, it is, thus, implied that its main targets are fellow members of the applied crystallography community. This paper should, however, also be of interest to the computational symmetry community because the underlying mathematical and statistical frameworks are identical when images are considered as *data planes* from which geometric-structural information is to be extracted and classified, regardless of the instruments with which they were recorded.

For the computational symmetry community [1,2,4,5] and with regards to Kanatani's associated developments in the robotics/computer vision fields [3,8-11], it is entirely natural to consider images as data planes. While this is largely because there are no microscopes and specifics of the underlying physics of the imaging process involved that may need modeling, modern microscopes are so good now that the data plane approach also works well in materials science and structural biology.

It should, therefore, not come as a surprise that this review follows the existing leads from the computational symmetry community but also goes beyond the current state of affairs in crystallographic symmetry classification schemes when multi-model inferences are discussed. The conclusion section of a recent review of the computational symmetry field states fittingly that *"strategies ... for handling real world complexity have to be developed to deal with ... the issue of subgroup relations among symmetry groups, raised by Kanatani"* [1]. The time seems indeed to be right for these kinds of developments and this paper reviews both the statistical foundation and the wider crystallographic implications of them. (The latter is mainly done in appendices, which may be of limited interest to members of the computational symmetry community.)

More or less 2D periodic Islamic building ornaments are assigned to plane symmetry groups in ref. [2] on the basis of the careful elucidation of the approximate site symmetries of conspicuous parts of periodic motifs in direct space. These elucidations rely, critically on *arbitrary thresholds* and thus they are always *subjective*. Their final plane symmetry group assignment can, therefore, not be objective.

Utilizing Kanatani's approach [3,8-11], the authors of ref. [2] could, in principle, transform their classifications to objective ones in spite of the multitude of "irregularities/defects" that their analyzed Islamic ornaments contain. When that was done, model selection uncertainties [12-15] would need to be addressed properly. A solution to the latter problem will be presented in this review as well. Note that model selection uncertainties are also not addressed in the work of Liu and coworkers [4,5] either.

The problems associated with the above mentioned subjective [1,2] crystallographic symmetry classifications, Kanatani's new statistical theory [3,8-11], and systematic ways of dealing with model selection uncertainties [12-15] became more relevant to the applied crystallography community with the recent emergence of both the crystalline "materials per design paradigm" [16] and model-based approaches to the imaging of crystals and long-range ordered materials.

References [17,18] utilize, for example, the above mentioned objective translation symmetry type classification scheme [6] for the detection and subsequent correction of double and multiple mini-tip artifacts in scanning tunneling microscope (STM) images of more or less 2D periodic arrays of molecules on a crystal surface by means of crystallographic image processing [19,20].

Independent of the type of microscope with which the data have been recorded, the purpose of crystallographic image processing is the extraction of geometric-structural information from noisy 2D periodic images. The translation and site/point symmetries in the hypothetical noise-free version of the image are taken advantage of as one averages over the asymmetric unit so that a better signal to noise ratio for the structure of interest is obtained. Note that the averaging over the asymmetric unit (rather than the translation periodic unit cell) ensures that better results are obtained than those achievable with traditional Fourier filtering [6]. This is because the multiplicity of the general position [21] boosts the number of entities over which one averages by a factor of up to 12.

The noisy 2D periodic image is considered to constitute a data plane and the models for the data at the foundation of crystallographic image processing are the 17 plane symmetry groups of 2D crystallography [21], which represent all possible combinations of translation and point/ site symmetries in the Euclidean plane. Crystallographic image processing originated about 50 years ago within the structural biology community [22] and contributed under the name "crystallographic electron microscopy" (monikers "Fourier or pseudo-kinematic electron microscopy") to the award of the 1982 Nobel Prize in Chemistry to Sir Aaron Klug.

Another type of model-based imaging in atomic resolution microscopy [23-27] with a complementary foundation originated as a very promising approach to quantitative transmission electron microscopy (TEM) at the University of Antwerp (Belgium) at the beginning of the $21^{st}$ century and led to the award of the 2017 Ernst Ruska Prize to Sandra Van Aert. The underlying procedures of that approach are analogous to single-crystal X-ray crystallography in so far as one distinguishes between the "solving" of the structure and the "refinement" of the resolved structure [23]. First the structure is resolved by the imaging of individual projected atomic columns in a more or less 2D periodic array with a state-of-the-art TEM. This is followed by a maximal likelihood refinement of the position and chemical composition of the atomic columns in that array.

Since the number of atoms in projected columns can be determined with single-atom accuracy when an aberration corrected TEM is utilized for the model-based imaging [26,27], a tomographic enhancement, i.e. the combing of structural information that was obtained from several atomic resolution images in different projections, was not necessary for the determination of the 3D structure of nanocrystals for which the thickness did not vary widely from atomic column to atomic column [26,27].

It is this author's opinion that the aforementioned model-based atomic-resolution approach to quantitative TEM could benefit from both the complementary geometric Akaike Information Criterion (G-AIC) approach that is outlined below in general terms and crystallographic image processing.

Reference [28] seems to be most suitable to illustrate the need for this review at the present time as it describes a geometric-structural feature extraction approach where a window is sliding over a noisy image of a crystal surface and the discrete window Fourier transform (dFT) is calculated at consecutive window positions of that atomically resolved image so that the locations of different crystal surface phases can be mapped in two dimensions. The authors of that paper state that it would in principle be possible to derive the local crystallography, i.e. the Bravais lattice type and plane symmetry group, of different types of more or less 2D periodic entities on crystal surfaces (or within crystalline matrices) from the data that they recorded with their sliding dFT windows in a scanning transmission electron microscope (STEM), but also caution that this "*would require substantial efforts at developing the appropriate image classification schemes*" [28].

Crystallographic classification schemes for 2D periodic patterns have been in existence for over nine decades [29], see ref. [21] for an authoritative, brief and mathematically comprehensive modern description as well as [30] for a good college level textbook. The real problem that needs to be addressed in the above mentioned context of the sliding dFTs is, however, *how* to make crystallographic classifications *objectively* on the basis of results from some *non-ideal* algorithm and when only *noisy* data are available, as it is the case in all real world applications.

The situation is analogous to what is encountered in the field of crystallographic 1D periodic classification schemes for gray-level patterns. The mathematical background of frieze symmetries and their projections from layer symmetries has been around for decades and is neatly summed up in an authoritative text [31], which follows the same outline as the comprehensive description of all plane symmetries of gray-level patterns [21] as projections from 3D space groups symmetries. The problem is again *how to* make classifications *objectively* on the basis of noisy experimental image data only without adding a subjective value judgment to arrive at one crystallographic symmetry class only.

More or less 1D periodic 2D images of crystalline materials such as aberration-corrected STEM images of plane coincidence site lattice (CSL) grain boundaries in edge-on projections which are atomically resolved [32-36] are known to be underlain by both predictable [37] types of frieze symmetries and 3D atomic level *bi*-crystal structures [38-41]. There is at present, however, no *objective* way to extract the parameters of grain boundary structures at the atomic level from such images. Subjectivity in the experimental determination of the very basic $\Sigma$ value (CSL index) has, for example, been recently discussed in ref. [42].

The core ideas of the crystallographic processing of noisy 2D images could be transferred to images that are periodic in 1D only as a first step towards the development of *objective* crystallographic symmetry classification schemes on the basis of Kanatani's statistical theory [3,8-11] and systematic ways of dealing with model selection uncertainties [12-15]. This would be equivalent to the adaptation of the proposal of this review to 1D periodic cases. The atomistic model-based approach that was pioneered at the University of Antwerp [23-27] could also be brought to bear on the extraction of geometric-structural information from atomic resolution images of grain boundaries. Appendix A and refs. [43-45] provide some more background on CSL (and approximate low-CSL index) grain boundaries in order to illustrate opportunities for 1D periodic symmetry classifications in that particular field.

As soon as suitable classification schemes have been demonstrated that work *without any arbitrarily set thresholds,* a robot can be programmed to classify input images automatically and sort them into crystallographic databases for more or less 2D or 1D periodic patterns *objectively*. It would then be to the user of such databases to (subjectively) interpret the objectively reported classification results. The author of this review presents here key aspects of his novel crystallographic symmetry classification scheme that is designed to work well in the presence of geometric-structural feature extraction uncertainties of the types that exist in more or less 2D and 1D periodic images.

Noise in the imaging process as well as geometric-structural feature extraction uncertainties in the processing of an image with some real world (non-ideal) algorithm will necessarily break all pre-existing symmetries of a crystalline sample (or that a synthetic image may possess due to its design) so that there will only be (non-genuine) pseudo-symmetries (of the second kind) left to be classified. The image is then, of necessity, only translation periodic to a larger or smaller extent so that none of the strict mathematically abstract restrictions of 2D [21] and 1D [31] crystallography are applicable anymore.

Further complications arise when there are *genuine* pseudo-symmetries [46] in the hypothetical noise-free version of a 2D or 1D periodic image. Geometric-structural feature extraction procedures can in the presence of noise not readily distinguish between non-genuine pseudo-symmetries that combine to form the underlying symmetry group structure of the hypothetical noise-free version of the image, on the one hand, and genuine pseudo-symmetries that exist in addition to this structure [46], on the other hand. Within this review, we will often refer to non-genuine pseudo-symmetries as pseudo-symmetries of the second kind. Appendix B provides more information on different types of pseudo-symmetries.

Because instances of the latter kind of pseudo-symmetries may be mistaken for instances of the former kind, the wrong underlying symmetry group structure may be inferred so that subsequent crystallographic classifications would be in error. Vice versa, due to noise in the experiments, non-genuine pseudo-symmetries may be mistaken for genuine pseudo-symmetries so that crystallographic symmetry classifications result which underreport the factual existing symmetry when an extrapolation to a zero-noise level is made. Genuine pseudo-symmetries also play important roles in twinning and the formation of multiple domains in crystalline solids [47].

Genuine pseudo-symmetry / genuine symmetry (pseudo-symmetries of the second kind) mix-ups that lead to symmetry classification problems in both inorganic crystal structures and molecule crystals (both small and large) in the presence of experimental noise are for the mainstream 3D crystallography case discussed in appendix C and refs. [47-74].

The three largest crystallographic databases for mainstream 3D crystallography results [75-79] are also briefly mentioned in appendix C1. Two of these databases are in open access [75,76,78,79]. References [80] and [81] concern specifics of single crystal X-ray protein crystallography.

A critical review of crystal structure determinations by means of single crystal X-ray crystallography in general is provided in ref. [82]. References [83-85] concern statistical descriptions in mainstream X-ray crystallography.

References [86-98] concern the preliminary re-analysis of the single crystal X-ray crystallography structure of a metal-organic framework compound that can probably be described as (*i*) incorrectly classified (due to an unrecognized pseudo-symmetry arising from the co-existence of triple domains) and (*ii*) incomplete due to the removal of electron density from the experimental results, see appendix C3. The crystallographic analysis of a few low electron dose STEM images, see ref. [95] or [94b] for one of these images, of that structure (as mentioned briefly in appendix C3) proved to be crucial to this author's arrival at this conclusion on that structure's validity [94a].

In spite of all of the kinds of difficulties that are mentioned above and because there seems to be an objective way for recognizing genuine pseudo-symmetries (in the presence of non-genuine pseudo-symmetries) as outlined below, it makes a lot of sense to assign a set of approximate crystallographic symmetry classifications to a 2D image so that the models one is using for atomic or molecular resolution imaging are of comparatively small dimensionalities and allow for optimal geometric-structural information extraction processes in the presence of noise.

Any real world geometric feature extraction algorithm will with necessity introduce some small systematic error into geometric-structural feature extraction results so that none of the computer programs that implement such algorithms will ever deliver *definitive* results [8,9,99] (Kanatani's dictum). Because there are no definitive feature extraction results, one should not attempt to classify these results into qualitatively exclusive (definitive) classes such as a single Bravais lattice type [6,21], Laue class [21b], and plane symmetry group [21] in the 2D case but utilize Kanatani's new statistics [8-11] instead.

This is because the traditional kinds of classifications imply that the extracted pseudo-symmetries adhere 100 % (i.e. *definitively*) to the restrictions that are imposed by a mathematically abstract crystallographic type, class, or group, which are all of a qualitatively strict nature per definition. Such an adherence can obviously not be genuine as there is noise in all image recording and processing steps in all *real world* applications.

In spite of this, allegedly definitive symmetry classifications are so far the common practice in both the computational symmetry and applied crystallography communities alike. They are, however, fundamentally unsound because all qualitative classifications will be in error insofar as they claim to be *definitive*, see Kanatani's comments from the year 1997 in this context [3].

Fortunately, crystallographic symmetries are hierarchic and the majority of them are non-disjoint [6,21,31]. These features allow for a boot-strapping approach that does not require an initial estimate of the generalized noise level in a more or less 2D or 1D periodic image.

By means of pair-wise comparison of non-disjoint models with Kanatani's G-AIC [8-11], one first obtains the model that minimizes the expected Kullback-Leibler information loss [12-15] within a set of models that represents a symmetry hierarchy branch and later on determines for this particular model the generalized noise level. When this has been achieved, one can calculate the relative likelihood that a model in a set of non-disjoint (or disjoint) models minimizes the expected Kullback-Leibler information loss and formulate so-called Akaike weights as conditional model probabilities that add up to 100 % for the whole set [12-15].

Instead of a definitive classification that makes a 100 % assignment to only one class, which cannot be guarantied to be correct due to the unavoidable presence of experimental noise and feature extraction uncertainties that are due to the utilized algorithm as discussed above, one obtains by this route a fuzzy classification that is spread over several classes of non-disjoint models within one symmetry hierarchy branch. One may also end up with a fuzzy classification that is spread over several classes of both non-disjoint and disjoint models if there is a genuine pseudo-symmetry [46] in the data plane.

The derived percentages of the adherences to the individual classes of models within (and outside of) a symmetry hierarchy branch will be specific to the noise level of the image to be classified and also very slightly specific to the algorithm with which the classification has been made. The effects of experimental noise and the utilized real world algorithm are summarized in a generalized noise level term. Reduced generalized noise levels of future image data from the same crystalline sample that are recorded with more sophisticated instruments and processed with more "truthful" feature extraction algorithms will have a tendency to change the individual percentages somewhat but will also never allow for definitive classifications. Also a reduction in the experimental noise level per unit cell can be obtained by the processing of a significantly larger image area that contains many more repeats of the 2D or 1D periodic motif.

Major goals of this review are to bring Kanatani's comments [3] and dictum [8,9,99] as well as his G-AIC approach [10,11] to the attention of both the applied crystallography and the computational symmetry communities. The utilization of the information theory concept of (*i*) Akaike weights [12-15] and (*ii*) their products [12] for complementing geometric-structural pieces of information (that were extracted from the results of the same imaging experiment or from the same synthetic data) for generalized noise-level dependent crystallographic classifications of more or less periodic crystal patterns constitute the novel ideas of this paper. This review will concentrate on crystal patterns in the form of 2D gray-level images that are more or less periodic in dimensions two and one.

Secondary goals of this review are popularizations of a Fourier space version of Liu's G-AIC for the assignments of plane symmetry groups to more or less 2D periodic images [4] and the author's versions of such criteria for Bravais lattice type [6] and Laue class assignments to such images. The combination of G-AICs for Bravais lattice types, Laue classes, and plane symmetry groups should be able to deal with the consequences of genuine pseudo-symmetries [46] that the hypothetical noise-free version of an image may possesses, either per design or by the nature of the crystalline sample from which it was recorded.

The rest of the paper is organized as follows. We begin with explaining the nature of Kanatani's comments on symmetry as a continuous and hierarchic feature in section 2.

This is followed by a discussion of Kanatani's dictum in section 3. Within that section, we will concern ourselves with genuine pseudo-symmetries [46], see Figure 1, which exist per design of both images and quote the related lattice parameter extraction results for these two images by three different algorithms/computer programs from ref. [99]. That part of this review is with necessity quite succinct as its only purpose is to illustrate the non-definitiveness of geometric-structural feature extraction results that are obtained by any real world algorithm from noise-free and noisy images alike.

Readers interested in the details of the three computer programs that implement these algorithms are referred to ref. [99] for comprehensive information. Two of these programs [100,101] are used in the applied crystallography community and the third [102] one supports all aspects of crystallographic image processing and electron crystallography on the basis of high-resolution (phase-contrast) transmission electron microscope (HRTEM) images [103] that were recorded within the validity range of the weak phase object (WPO) approximation. While one of these programs [100] and most algorithms of the computational symmetry community [1] work in direct space, the other two programs [101,102] that were utilized in ref. [99] work in Fourier/reciprocal space.

For easy references below, we will use the capital letters A, B, and C instead of either the actual names of these three computer programs or their entries in the final list of references at the end of this paper. Table 1 provides the conversion key.

**Table 1.** Letter key for references to the three algorithms/ computer programs for which we will quote results in this paper that were taken from ref. [99].

| Algorithm's number in the final reference section | Algorithm's letter reference in this paper |
|---|---|
| [100] | A |
| [101] | B |
| [102] | C |

The 4th section on G-AICs reviews first the general form of these criteria and then proceeds by giving specifics of Fourier space versions of such criteria for fuzzy, i.e. quantitative generalized noise-level dependent, classifications of geometric-structural feature extraction results into plane symmetry groups, Laue classes, and Bravais lattice types. Liu's and her co-workers' frieze pattern assignments to time series recording of both a walking humanoid avatar and a walking human being [4] will be mentioned in this section briefly (and discussed further in appendix D) as illustrations of the fact that one should not only report the most likely crystallographic symmetry classification for a real world experiment, but also its relative likelihood as well as the likelihoods of reasonable alternatives in order to make a fair assessment of the crystallographic model selection uncertainty [12-15].

In the 5th section, we will provide equations for the relative likelihoods of disjoint and non-disjoint crystallographic symmetry models within a set, their respective mutual evidence ratios, and their Akaike weights. There are also equations for the usage of Akaike weights for multi-model predictions that are based on the relative probabilities of crystallographic symmetry models within a set. Section 5 ends with the equations for combined posterior model probabilities [12] that are based on complementing pieces of geometric-structural information in more or less 2D periodic (noisy) images.

The corresponding combined Akaike weights should be helpful for distinguishing between genuine pseudo-symmetries [46] and non-genuine pseudo-symmetries that the hypothetical noise-free version of an image processes.

The 4th and 5th sections constitute the core of this review and contain the equations that refine its novel ideas. Finally, there is a brief summary and conclusions section.

As already mentioned above, there are four appendices that present: (*i*) the potential of the main proposal of this review with respect of the extraction of grain boundary structures from atomic resolution images that are more or less periodic in 1D, (*ii*) different types of pseudo-symmetries in general terms, (*iii*) pseudo-symmetry mediated mis-classifications in both the scientific literature and the major databases of mainstream 3D crystallography as well as a brief discussion of the crystallographic *R* value, and (*iv*) crystallographic comments on the only so far existing experimental 1D periodic study that utilized a geometric Akaike Information Criterion.

## 2. Kanatani's comments on symmetry as a continuous and hierarchic feature

At the core of Kanatani's comments is the observation that symmetries must with geometric necessity be part of disjoint and non-disjoint hierarchy branches. This applies obviously to both crystallographic and non-crystallographic symmetries alike, although no such distinctions were made in ref. [3] as a few non-disjoint and disjoint point symmetries were discussed exclusively.

Because hierarchy branches do exist, crystallographic symmetry types, classes, and groups are quite often non-disjoint. For example, a hexagonal rhombus (with an angle of 120° between its two edges of equal length) is higher up in the symmetry hierarchy than a general rhombus (where this angle is neither 120° nor 90°). The general rhombus, on the other hand, is higher up in the hierarchy than a parallelogram where the two edges have different lengths and the angle is not 90°.

The square and hexagonal rhombi are, however, disjoint as they are at the top of different hierarchy branches of the quadrilaterals that serve as crystallographic unit cells [6,98]. Analogously, the rectangle and the general rhombus are disjoint as they are part of different hierarchy branches.

Kanatani illustrates in ref. [3] that a consequence of these kinds of hierarchies is that one can never assign extracted geometric-structural features in an objective way to a more constrained symmetry type on the basis of a distance measure *alone*.

## 3. Kanatani's dictum

A direct quote from refs. [8,9] is in order here to start this section: *"The reason why there exist so many feature extraction algorithms, none of them being definitive, is that they are aiming at an intrinsically impossible task."* While this statement might be somewhat shocking to researchers who never before thought about this topic deeply, it is certainly true. No real world feature extraction algorithm working on real world data will ever be able to deliver *definitive* results. This is because all algorithms (and the computer programs that implement them) are based on heuristics and use approximations as well as internal thresholds to achieve their goals. Also all real world image data are of finite resolution and noisy.

As mentioned above, a thorough illustration of Kanatani's dictum within a crystallographic context is provided in ref. [99]. We take from that paper the lattice parameter extraction results from the two images that are shown in Figure 1.

Results that were obtained in the default settings of three different computer programs that implement three different types of algorithms (A to C in Table 1) are listed in Tables 2a and 3a. Tables 2b and 3b list, on the other hand, re-interpreted/re-calculated results from algorithm B (on the basis of the displayed dFT amplitude maps) and results that were obtained in a non-default setting of algorithm C.

The two images in Figure 1 are synthetic and freely downloadable (together with many more images of the same size and type) at the website that is listed as ref. [104]. On the left hand side of this figure, there is the noise-free (original) image of the pair. The image on the right hand side of this figure has been obtained by adding independent Gaussian noise of mean zero and a standard deviation of 10 % of the maximal image intensity to the individual pixels of the noise-free image to the left.

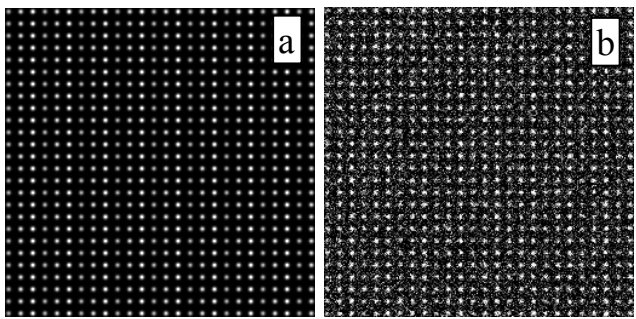

**Figure 1. (a)** Image with plane symmetry group *pm* that possesses genuine pseudo-symmetries per design which are in **(b)** exacerbated by added independent Gaussian noise of mean zero and a standard deviation of 10 % of the maximal image intensity. The lattice in (a) is visibly of the rectangular type [6]. In the noisy image, the lattice is apparently of the square type. Both images are in open access [104] and are reproduced here with CC-BY (share — copy and redistribute the material in any medium or format adapt — remix, transform, and build upon the material for any purpose, even commercially) licenses. The labeling of the images with the letters *a* and *b* are the only modifications that were made.

Note that Kanatani includes per definition all kinds of image feature extraction uncertainties into the generalized noise term in his G-AICs so that one cannot extract definitive results even from the image in Figure 1a, which is free of added Gaussian noise. In this review, noise is treated in the generalized sense that is in accord with Kanatani's dictum [8,9].

Both images in Figure 1 feature genuine pseudo-symmetries [46]. The Bravais lattice of these images is per design of the rectangular (primitive) type with an a/b ratio of three to one. The asymmetric unit consists of three vertical blobs with different (but somewhat similar) intensities and sizes. It is the apparent similarity in the intensity and size of these blobs that causes a translational pseudo-symmetry.

A motif-based (four-fold rotation plus mirror lines) pseudo-symmetry is also present in both images due to the 3 to 1 ratio of the lattice constant magnitudes parallel to the horizontal and vertical edges of the images. The four-fold rotation plus mirror lines pseudo-symmetry contains in itself a two-fold rotation plus mirror lines pseudo-symmetry. The added noise exacerbates in Figure 1b the pseudo-symmetries that are already visible in Figure 1a.

As already mentioned above, the main thrust of ref. [99] was to illustrate Kanatani's dictum on multiple examples. Since three algorithms/computer programs were applied to a total of 12 images in [99], a measure of the reliability of subsequent geometric inferences on the basis of the outputs of the computer programs that implemented these algorithms was also obtained.

Because the three algorithms were tested on both noise-free images (such as the one shown in Figure 1a) and noisy images that were derived from the noise-free images (such as the one in Figure 1b), the robustness of the algorithms/computer programs in the presence of Gaussian noise was also tested in [99].

Somewhat surprisingly, Table 2a shows that only one of the three tested algorithms extracted qualitatively correct lattice parameters from the noise-free, but visibly pseudo-symmetric, image in Figure 1a. These lattice parameters are in good compliance with the rectangular Bravais lattice type that this image possesses per design. For easy reference, qualitatively correct results are marked in bold font in all of the four image data tables in this review.

Only algorithm A was, thus, capable of dealing with the translational pseudo-symmetry in Figure 1a effectively as its lattice parameter extraction results are given in bold font in Table 2a. The other two algorithms extracted in their default settings a unit cell that is too small by a factor of three from this figure. This is also reflected by the ratio of the two basis vectors, which was incorrectly determined as nearly unity by algorithms B and C in their respective default settings [99].

**Table 2a.** Extracted lattice parameters from the noise-free image in Figure 1a and derived unit cell area utilizing the default settings of the three programs. The qualitatively correct result is marked in bold font.

| Algorithm's ref. letter | b / a | γ in ° | Unit cell area in square pixels |
|---|---|---|---|
| A | **0.333 ± 0.06** | **90.0 ± 0.4** | **300.0 ± 2.0** |
| B | 1.004 ± 0.01 | 90.0 ± 0.05 | 99.4 ± 1.0 |
| C | 0.998 ± 0.01 | 90.0 ± 0.05 | 99.8 ± 1.0 |

Extracted basis vectors that nearly possess the same magnitude and are also perpendicular to each other within error bars are of course what one would expect for a square Bravais lattice. In other words, to the algorithms B and C in their default settings, the existing (genuine) translational pseudo-symmetry [46] in Figure 1a was apparently a genuine crystallographic symmetry since "quantitatively wrong" lattice parameter sets were extracted.

It was straightforward to re-interpret/re-calculate the lattice parameter extraction output for Figure 1a as obtained with algorithm B on the basis of the dFT amplitude map that the program displayed [99]. This resulted in a bold font entry for qualitative correctness in Table 2b for algorithm B. For algorithm C, using a non-default setting in the processing of Figure 1a also resulted in a bold font entry in this table.

**Table 2b.** Extracted lattice parameters from the noise-free image in Figure 1a and derived unit cell area after a re-interpretation of the results from algorithm B and as obtained in a non-default setting of algorithm C. Both results are qualitatively correct and, therefore, marked in bold font.

| Algorithm's ref. letter | b / a | γ in ° | Unit cell area in square pixels |
|---|---|---|---|
| B | 0.335 ± 0.01 | 90.0 ± 0.05 | 298.2 ± 3.0 |
| C | 0.333 ± 0.02 | 90.0 ± 0.05 | 300.0 ± 2.0 |

The added noise in Figure 1b "fooled" all three computer programs (in their default settings) into extracting results that are obviously incorrect, see Table 3a. This is a direct consequence of the noise-exacerbated translational pseudo-symmetry in the image shown in Figure 1b.

**Table 3a.** Extracted lattice parameters from the noisy image in Figure 1b and derived unit cell area. There is no qualitatively correct result to be marked in bold font.

| Algorithm's ref. letter | b / a | γ in ° | Unit cell area in square pixels |
|---|---|---|---|
| A | 1.414 ± 0.07 | 134.2 ± 0.6 | 98.6 ± 1.5 |
| B | 0.995 ± 0.01 | 89.8 ± 0.05 | 99.1 ± 1.0 |
| C | 1.000 ± 0.01 | 90.0 ± 0.05 | 100.0 ± 1.0 |

The oblique unit cell that algorithm A extracted from the image in Figure 1b, see Table 3a, can be straightforwardly transformed into a pseudo-square unit cell with essentially the same parameters as those that were obtained with the other two algorithms. The corresponding transformation matrix from the old (unprimed, oblique) lattice vectors to the primed (new, pseudo-square) lattice vectors is given by $\begin{pmatrix} 1 & -1 \\ 0 & 1 \end{pmatrix}$ since $\vec{a}' = \vec{a}$ , $\vec{b}' = -\vec{a} + \vec{b}$ , $\gamma' \approx 90°$, with $|\vec{b}'| \approx |\vec{a}'|$ and $|\vec{b}| \approx \sqrt{2}|\vec{a}|$, $\gamma \approx 90° + 45°$.

As the determinant of this matrix is +1, both unit cells represent the same area and their right handedness is preserved. (The information that the correct unit cell is actually of the rectangular type and three times larger than obtained with algorithm A can obviously not be recovered from the corresponding entry for this image in Table 3a.)

Re-interpreting/re-calculating the lattice parameter extraction outputs for Figure 1b as obtained with algorithm B (on the basis of the dFT amplitude map of that image) and using a non-default setting of algorithm C in the processing of this image led to qualitatively correct results and bold font entries for both algorithms in Table 3b.

**Table 3b.** Extracted lattice parameters from the noisy image in Figure 1b and derived unit cell area after a re-interpretation of the results from algorithm B and as obtained in a non-default setting of algorithm C. Both results are qualitatively correct and, therefore, marked in bold font.

| Algorithm's ref. letter | b / a | γ in ° | Unit cell area in square pixels |
|---|---|---|---|
| B | 0.332 ± 0.01 | 89.8 ± 0.05 | 297.3 ± 3.0 |
| C | 0.333 ± 0.02 | 90.0 ± 0.05 | 300.0 ± 2.0 |

The stated error bars on the unit cell angles of 0.05° for the two algorithms/computer programs that extract lattice parameters in Fourier space, i.e. B and C, are based on the implied number of significant figures output by one of these programs [99], but seem to be too small to allow for agreement of the extraction results of the different algorithms in the case of lattice parameter extractions by the default program settings from the noisy image in Figure 1b.

The traditional way of assigning Bravais lattice types to the lattice parameters of the two images in Figure 1 that have been extracted by three different algorithms within the stated error bars as listed in Tables 2a,b and 3a,b may, obviously, lead to mis-classifications given the numerical variations in Tables 2a to 3b. If one does not know the design parameters and history of the two images in Figure 1 in advance, one is hard pressed to figure out which of the results in these four tables are actually trustworthy, let alone to make *definitive* classifications into Bravais lattice types. One would certainly be ill advised to average the results from the three different algorithms in Tables 2a and 3a.

Guided by the "somewhat squarish" visual appearance of what appears to be unit cells in the image of Figure 1b, most researchers would probably classify that image as belonging to the square Bravais lattice type. Two of the results listed in Table 3a would support this classification in the traditional way based on the numerical values of the extracted lattice parameters and their somewhat extended error bars. This would, however, be incorrect!

A *fuzzy* classification into Bravais lattice types on the basis of translation symmetry model probabilities (Akaike weights) would, on the other hand, be noise-level dependent and correct in a fundamental sense. Likewise fuzzy classifications into (*i*) Laue classes on the basis of point symmetry model probabilities and (*ii*) plane symmetry groups on the basis of plane symmetry model probabilities, both utilizing complementing types of Akaike weights, would also be correct in a fundamental sense and generalized-noise level dependent.

The crystallographic symmetry classifications of the image in Figure 1a would obviously be much less fuzzy than those of the image in Figure 1b, although still not completely definitive as a matter of principle when a real world algorithm is involved. It is expected that the crystallographic classifications of both images would peak for plane symmetry group *pm*, Laue class *2mm,* and the rectangular (primitive) Bravais lattice type. This is because these crystallographic categories went into the design of both images.

In case of the image in Figure 1a, the peaking at these crystallographic categories will be much sharper than for the image in Figure 1b because only geometric-structural feature extraction uncertainties that are due to the particulars of the applied algorithms/computer programs will make the classifications of the former image fuzzy (as there is no added Gaussian noise present that disturbs the recognition of the design categories).

## 4. Geometric Akaike Information Criteria (G-AICs)

### 4.1. General considerations

All G-AICs transfer the central idea of the very widely employed Akaike Information Criterion (AIC) [105,106] of traditional statistics, which is based on the asymptotic limit of an infinite number of observations, to Kanatani's new type of statistics where a vanishing noise level serves as asymptotic variable and where there is typically only one observation/image [8-11].

The following direct quote from Hirotugu Akaike's original paper [105]: *"AIC = (-2) log (maximum likelihood) + 2 (number of independently adjusted parameters within the model)"* illustrates that the "accuracy" of a model which constitutes the first term (and is obtained from a maximal likelihood estimation of the model's parameters) is balanced by the "complexity" of the model (by means of a penalty for having a certain number of free parameters available for the fitting of the model to the data), which constitutes the second term. The *"log"* in this quote refers to a logarithm to the basis of Euler's number.

More precisely, the negative log-likelihood score of a model is a measure of the lack of its fit to the data. It forms the first term in an AIC. The second term of an AIC is simply a penalty for greater model complexity, i.e. represents a bias correction. When two frequency based models for the same data are compared with respect to their predictive power, the model that possesses the smaller AIC value is considered to be the better one.

Traditional AICs as stated above and versions of them that account for unfavorable ratios between the number of observation and the number of model parameters [12] are very widely used in countless branches of science and engineering [13-15] as for example attested to by the very large number of citations to Akaike's original paper. (For that paper's Google Scholar citation count at the time of finalizing this review, see the entry for [105] in the list of references).

Information criteria that are either derived from Akaike's AIC or are based on traditional (frequentist) statistics alternatives to this criterion are also at the core of the quantitative model-based atomic resolution TEM approach [24] that was mentioned in the Introduction and Background section.

Akaike referred to his criterion simply as *"an information criterion"* [105]. The acronym MAICE, which stands for *"minimum information theoretic criterion (AIC) estimate"*, was also introduced by him. Akaike wrote in 1974 that *"the need of the subjective judgment required in the hypothesis testing procedure ... is completely eliminated"* by the utilization of MAICEs because *"the problem of statistical identification is explicitly formulated as a problem of estimation"* [105].

As already mentioned above, geometric AICs contain two terms that are analogous to the two terms in the traditional (frequentist) AIC. The particulars of the form of the second term depend on the types of geometric models in a set, from which the one that minimizes the G-AIC value is to be selected as the best model for representing the image data information.

For the practical application of a G-AIC, noise in images must be (to a sufficient approximation) of the "white" Gaussian type and systematic errors in the imaging and algorithmic processing procedures must be small in comparison to random errors. Because G-AICs are first order approximations, the generalized noise level must be reasonably small. All these preconditions are fulfilled by image recordings with certain modern scientific instruments, where the extraction of model parameters from the recorded data planes proceeds to a large extent independently of the particulars and type of the instruments, e.g. STMs, STEMs, …, HRTEMs (in the WPO approximation) as mentioned in the Introduction and Background section, with which image data has been recorded [6,17-20,22-28,100-103].

With independent to a large extent, this author means that specifics of the point-spread function of a microscope should be included when better results are required. Approximate results are to be expected when these specifics are ignored. Similarly, the real structure of crystalline samples could either be included for better results or ignored for approximate results.

By virtue of the central limit theorem of frequentist statistics, the white (Gaussian) noise requirement of G-AICs can often be considered as fulfilled when there are several different types of noise in a microscopical imaging process [23]. Different types of noise will originate from different sources, but none of these sources is allowed to be dominant for this theorem to be valid.

As a very notable difference to the AIC of traditional statistics, noise is not a model parameter in Kanatani's G-AIC. Vanishing noise is instead the asymptotic limit in this new kind of statistics in a way similar to the number of observations going to infinity in frequentist statistics. The requirement of Gaussian noise results in maximal likelihood determination procedures of the model parameters that take on the form of least squares fits of the models to the data in G-AICs. Also the number of data points and the dimension of the model enter the penalty term of the equation of the G-AIC, while their counterparts are absent in the equation for the traditional AIC.

When the experimental noise in a more or less 2D periodic image is of the Gaussian type, the standard deviations of the mean values of the intensity of corresponding pixels in repeating unit cells decrease with the square root of the number of repeats. This allows for more precise estimates of the mean values of the intensity of the group of individual pixels that collectively form the 2D periodic unit cell. Loosely speaking, this is analogous to a reduction of the noise level per unit cell. The disturbing effects of the 10 % Gaussian noise in the image in Figure 1b are thereby reduced to less than 1 % per unit cell.

## 4.2. G-AIC for plane and frieze symmetry groups

Liu and coworkers state in [4a] (and the appendix of the earlier version of their paper [4b]) that the dimension of their data space is one and the dimension of their model space is zero, so that a co-dimension of one results. Simple algebra leads to the following ratio of the least-square residual, $J$, of a more symmetric model, $S_{more}$, to the least-squares residual of a less symmetric (more general) model ($S_{less}$) that is non-disjoint:

$$\frac{J[S_{more}]}{J[S_{less}]} < 1 + \frac{2(k_{more} - k_{less})}{k_{more}(k_{less} - 1)} \qquad (1),$$

which allows one to conclude that the (non-disjoint) more symmetric model is the one of the two models that minimizes the expected Kullback-Leibler information loss when one deals with the unit cell content. The variable $k$ is thereby the so called multiplicity of the general position of a plane [21] or frieze [31] symmetry group. In the language of 1D and 2D crystallography, the non-disjoint relations of set theory are referred to as maximal non-isomorphic translationengleiche and klassengleiche type IIa subgroup-supergroup relationships, as exhaustively tabulated in refs. [31] and [21].

Note that although the least squares residual of the more general (less symmetric) model will typically be smaller than its counterpart for the less general (more symmetric) model or at most equal to it, equation (1) still allows for the selection of the more symmetric model if its residual is not too large. This is because the model accuracy is *balanced* in the G-AIC by a penalty term that includes both the higher multiplicity of the general position of the more symmetric model and the lower multiplicity of the general position of the less symmetric model.

There was only one application of equation (1) to the identification of most likely frieze symmetry groups [4,5] so far in the scientific literature and none to the identification of most likely plane symmetry groups of which the author of this review is aware. This state of affairs might be due to the necessary computational effort for obtaining the residuals in direct space as the sums of all squared differences of pixel intensities between the raw image and its symmetrized versions [4,5]. Also, there is the non-trivial issue of aligning the raw image and its symmetrized versions in direct space, see appendix D for possible consequences of misalignments and ignorance of the crystallographic origin conventions [21,31].

There is, however, a straightforward way to overcome both of these problems in Fourier space that goes by the name of crystallographic image processing [19,20,22]. Origin alignments are actually straightforward in reciprocal space and part of the crystallographic image processing procedures.

Because the intensity values of all pixels contribute to all Fourier coefficients (FCs), the sums of the squared differences of the complex FCs of the raw image and its symmetrized versions can be calculated in Fourier space and be substituted into equation (1) for the real space residuals. The complex FCs residuals enter equation (1) without modification of its right hand side because one can substitute the multiplicity of the general position in direct space for the number of symmetry operators of a plane symmetry group. This number is the same in both direct [21] and reciprocal/Fourier space [107]. The Fourier space approach to the interpretation of equation (1) is enabled by Fourier coefficient model residuals that approximate [108] direct space least-squares model fitting residuals sufficiently well.

While a more or less 2D periodic square image with an edge length of 512 pixels possesses 262,144 individual pixels, there may be less than a hundred FCs to represent all of the information that is contained in it in Fourier space. A large reduction in the computational effort will, therefore, result when one works in Fourier space for the determination of the ratios of the least-squares residuals in equation (1).

The precondition for going into Fourier space is from this author's experience that there are at least some 50 – or better yet more than 100 – unit cell repeats in the image in order to keep series truncation artifacts and edge effects small. Equation (1) implies the presence of an integral number of unit cells in the more or less 2D periodic image. Artifacts due to incomplete unit cells and the multiplicities of special positions [21,31] become negligible when the pixels are numerous and there are many repeats of the more or less translation periodic motif in the image.

While large numbers of repeating noisy unit cells are often not present in images that are usually studied by the computational symmetry community [1,2,4,5], they are commonplace in images that are processed by the applied crystallography community [6,17-20,22-28,100-103].

## 4.3. G-AIC for 2D Laue classes

The Laue class of a more or less 2D periodic image is visibly displayed in the amplitude map of its dFT. As a matter of fact, the FC amplitudes are laid out in such a map as discrete values at the positions of the nodes of the reciprocal lattice of the image.

One can, therefore, use equation (1) as well when one bases the residuals on the sums of the squared differences between the dFT amplitudes of the raw image and the dFT amplitudes of its symmetrized versions. Crystallographic image processing [19,20] provides again the means to obtain the residuals in a computationally efficient manner. The same kinds of considerations of the number of pixels in the image and the number of repeating noisy unit cells in direct space apply as in the previous section.

There are six Laue classes in 2D in both direct and reciprocal/Fourier space. In the latter space, the Laue symmetry classes are defined with respect to the central 0,0 FC amplitude peak in a dFT.

When the Laue class is, for example, assigned to the image in Figure 1a by means of a G-AIC, one will for sure obtain the highest model probability for class *2mm*. This is because the plane symmetry of that noise-free image is per design *pm*, which is "mathematically linked" to Laue class *2mm*. The visible four-fold rotation plus mirror lines (motif-based) pseudo-symmetry in this figure will then be revealed as such.

Plane symmetry *pm* is of the non-centrosymmetric type [21], which means the phase angles of the FCs of the image intensity are not all restricted to be either 0° or 180°. This fact should come in handy when one is trying to deal with the noise-exacerbated pseudo-symmetries in Figure 1b. The associated Akaike weight products of *joint* fuzzy assignments to Bravais lattice types, Laue classes, and plane symmetry groups (on the basis of the applicable G-AICs) should be able to reveal the pseudo-symmetries also for this image.

One should, therefore, at least for manifestly pseudo-symmetric images, strive for combinations of G-AICs and Akaike weight products in order to make the best use of the available types of complementing geometric-structural information in more or less 2D periodic images.

### 4.4. G-AIC for 2D Bravais lattice types

Reference [6] describes the G-AIC for Bravais lattice type assignments to lattice parameters that were extracted from more or less 2D periodic images. For identifying the (non-disjoint) higher symmetric translation symmetry model that minimizes the expected Kullback-Leibler information loss, the following inequality:

$$\frac{J[S_{more}]}{J[S_{less}]} < \frac{2L_{more} - L_{less}}{L_{less}} \qquad (2)$$

suffices, where $L$ is the number of constraints on a quadrilateral that serves as the shape of a crystallographic unit cell. Table 4 lists these numbers for easy reference.

**Table 4.** Number of constraints that enter equation (2) in a G-AIC for the fuzzy classification into Bravais lattice types.

| parallelogram | rectangle | general rhombus | square | hexagonal rhombus |
|---|---|---|---|---|
| 2 | 3 | 3 | 4 | 4 |

### 4.5. When is a noise level estimate mandatory?

There is obviously no need to make an estimate of the generalized noise level of a more or less 2D periodic image in order to use equations (1) and (2). This is due to our dealing with non-disjoint models in these cases, i.e. with models within a crystallographic symmetry hierarchy branch. One can for the comparison of a pair of such models, as implicitly stated in equations (1) and (2), eliminate the need to know the noise level by algebraic means.

As already mentioned above, comparing disjoint models does, on the other hand, require an estimate of the generalized noise level (equation 3 below) on the basis of the best model in the set. Such estimates are of particular importance when there are both genuine [46] and non-genuine pseudo-symmetries in a noisy 2D periodic image that is to be classified.

## 5. Utilizing Geometric Akaike Information Criteria

### 5.1. Highlights of the underlying information theory

When one has identified the model that minimizes the expected Kullback-Leibler information loss, i.e. the so-called "Kullback-Leibler best" [12-15] model, by a boot-strapping approach within a symmetry branch, one can obtain a good estimate for the noise level on the basis of this model by the following equation:

$$\hat{\sigma}^2 \approx \frac{J[S_{best}]}{r_{best}N - n_{best}} \qquad (3),$$

where $r$ is the co-dimension, $N$ is the number of data points, $n$ is the degree of freedom, and the subscript $_{best}$ stands for the best model in the set [9-11]. (The hat over the sigma means that it is an estimator.)

For this particular model, one can be confident to have extracted the maximum of geometric-structural information from the image by the least-squares fitting of the model parameters to the image data while also separating out the "non-information" that is summed up in the generalized noise estimate of equation (3).

The estimate of the noise level in the image according to equation (3) becomes part of the full (first-order) equation for the G-AIC of all $i$ models $S_i$ within a set:

$$AIC_i[S_i] = J_i[S_i] + 2(d_iN + n_i)\hat{\sigma}^2 \qquad (4),$$

where $d$ is the dimension of the model.

All G-AIC values are relative and on the scale of information. Only the relative differences of the $AIC_i$ values of the $i$ models in either a disjoint or a non-disjoint set matter for crystallographic symmetry classifications.

These differences are standardized on the basis of the Kullback-Leibler best model in the set, i.e. the one for which we obtained the minimum G-AIC value from equation (4). The (standardized or rescaled) model specific G-AIC differences are obtained by:

$$\Delta_i = AIC_i - AIC_{best} \qquad (5),$$

for all $i$ models in a set of models. This kind of difference is obviously zero for the best model in the set, which possesses the smallest G-AIC value and is designated by the subscript $_{best}$.

The relative likelihoods of all $i$ models in the set are obtained by Akaike's transformation [12-15,106]:

$$\ell_i \propto \exp(-\tfrac{1}{2}\Delta_i) \qquad (6),$$

whereby $\propto$ stands for "proportional to" and the pre-factor of ½ on the standardized G-AIC differences is due to the original definition of the traditional AIC [105,106]. The best model in the set obtains a relative likelihood of unity and all other models come in at a fraction of unity.

The evidence ratio of one model with respect to another one within the model set is obtained by:

$$E_{i,j} = \ell_i \Big/ \ell_j = \exp(-\tfrac{1}{2}\{\Delta_i - \Delta_j\}) \qquad (7).$$

Evidence ratios have a *"raffle ticket interpretation"* in quantifying the strength of evidence in favor of one model with respect to another in the same set [12]. When, for example, the evidence ratio of model X with respect to model Y is 20, there is at the very least moderate if not strong evidence in support of model X. The difference of the two related relative G-AIC differences in the inner parentheses in equation (7) is then approximately 6.

This is analogous to model X possessing 20 raffle tickets while model Y possesses only a single ticket. Clearly model X is then more likely to win a raffle and the evidence in support of it is stronger than for model Y, which is, however, not to be discarded as it is not entirely without merit given the generalized noise level [12].

Note in passing that twenty to one odds are incidentally also the basis of many traditional ad hoc "tail probability threshold ≤ 0.05" null hypothesis ("P-value significance, α-level") testing schemes [13,14]. (Walter Clark Hamilton pioneered the application of such a hypothesis testing scheme for nested crystal structure models in mainstream 3D crystallography on the basis of the ratio of generalized crystallographic "*R* factors" in the year 1965 [85].) The information theory based approach utilized above is, however, much more powerful [12-15] than that kind of traditional hypothesis testing and does not "clip off" models with moderate and small likelihoods as they could turn out to be correct when data with significantly reduced noise levels becomes available in the future. In Kenneth P. Burnham's and David R. Anderson's own words: *"information-theoretic criteria ... are not a 'test' in any sense, and there are no associated concepts such as test power or P-values or α-levels"* [14].

Individual model probabilities that add up to 100 % for the whole set of *R* models are commonly referred to as either Akaike weights [12-15] or Bayesian posterior model probabilities [12]. These probabilities are obtained by the normalization of the relative model likelihoods:

$$w_i = \frac{\exp(-\tfrac{1}{2}\Delta_i)}{\sum_{r=1}^{R} \exp(-\tfrac{1}{2}\Delta_r)} \cdot 100\% \qquad (8),$$

whereby a given $w_i$ is the probability that model *i* is the Kullback-Leibler best model. Akaike weights for a subset of models are additive and can be summed into confidence sets [12]. (Obviously, the sum of all Akaike weights is 100 %). While summing into confidence sets is somewhat subjective, there is certainly no arbitrariness in the usage of the equations of this review.

Akaike weights are also useful for the averaging of model parameters and predictions that are based on a multitude of "low-$\Delta_i$ models" within a set [12-15]. Model parameters are in the context of this review the values of the unit cell parameters in direct and reciprocal space, the discrete Fourier coefficient amplitude and phase angels of the image intensity that form (in reciprocal space) the "Fourier equivalent" [107] of the asymmetric unit of a plane symmetry group, and the gray level values of the group of individual pixels that collectively form the asymmetric unit in direct space.

Higher symmetric translation symmetry, Laue symmetry, and plane symmetry models possess obviously fewer parameters than their lower symmetric counterparts. Just as unit cell parameters are restricted by translation symmetries in all Bravais lattice types higher than oblique, the values of the gray levels of the individual pixels that form a unit cell in direct space are restricted by site symmetries higher than the identity rotation in all plane symmetry groups higher than *p1* and *pg*.

Typical predictions of higher symmetric plane symmetry models in direct space are the values of all pixels in the unit cell rather than just of those pixels that form collectively the asymmetric unit. Model predictions in Fourier space refer to the whole discrete and complex reciprocal data plane rather than just the Fourier space equivalent of the asymmetric unit [107].

Model averaged parameters or predictions are simply the weighted averages over parameters or predictions within a model set

$$\hat{\theta}_{average} = \sum_{i=1}^{R} w_i \cdot \hat{\theta}_i \qquad (9),$$

whereby the "hat" over the symbol for the parameter or prediction refers to an estimate [12]. An estimator of the variance of a parameter or prediction estimate that incorporates a variance component for model selection uncertainty is given by

$$\hat{var}(\hat{\theta}_{average}) = \sum_{i=1}^{R} w_i \cdot \left\{ \hat{var}(\hat{\theta}_i \mid g_i) + (\hat{\theta}_i - \hat{\theta}_{average})^2 \right\} \qquad (10),$$

whereby $g_i$ is the $i^{th}$ model and the extended notation $\hat{\theta}_i \mid g_i$ clarifies that the parameter or prediction estimator is in each of the *R* cases specific to a model in the set.

The variance estimate assesses the precision of the parameter or prediction estimate over the considered set of models and allows for the generation of confidence intervals that incorporate a measure for the model selection uncertainty. The standard error of a parameter or prediction is in multi-model averaging given by

$$se(\hat{\theta}_{average}) = \sqrt{var(\hat{\theta}_{average})} \qquad (11a),$$

so that a 95 % confidence interval [12] (or reasonable error bar widths on the model averaged parameter or prediction in other words) can be approximated by

$$\theta_{average} \pm 1.96 \cdot se \qquad (11b).$$

Equation (8) may also be utilized to combine for a more or less 2D periodic image the probability of its fuzzy classification into a Bravais lattice type with the probability of its fuzzy classification into a Laue class. The combined fuzzy classification of such an image into a Bravais lattice type and a plane symmetry group is also possible on the basis of equation (8). This particular combination of fuzzy classifications is probably effective for dealing with genuine pseudo-symmetries in the presence of noise and will be further discussed below in section 5.2 with reference to the images in Figure 1.

When a set of discrete prior probabilities on models, $p_q$, exists (in a Bayesian sense [12]) that is best at representing complementary (other) aspects of the finite information in the image data, one is justified to obtain "updated" Bayesian posterior model probabilities by an extension of equation (8) to:

$$w_{i,q}^{\ updated} = w_i \cdot w_q = \frac{\exp(-\frac{1}{2}\Delta_i) \cdot p_q}{\sum_{r=1}^{R} \exp(-\frac{1}{2}\Delta_r) \cdot \sum_{h=1}^{H} p_h} \cdot 100\% \quad (12),$$

whereby:

$$\frac{p_q}{\sum_{h=1}^{H} p_h} = w_q \tag{13}.$$

$$\frac{p_q}{\sum_{h=1}^{H} p_h} = \frac{\exp(-\frac{1}{2}\Delta_q^{\ other\_aspects})}{\sum_{h=1}^{H} \exp(-\frac{1}{2}\Delta_h^{\ other\_aspects})} \cdot 100\%$$

Fuzzy plane symmetry group and Laue class assignments complement fuzzy Bravais lattice type assignments in 2D because all three of them are based on different (but complementary) pieces of geometric-structural information in the same complex dFT data plane. Equation (12) may, therefore, be expanded by a third factor as defined by equation (13). A combined Akaike weight with factors for Bravais lattice types, Laue classes, and plane symmetry groups would constitute some kind of a comprehensive probabilistic crystallographic symmetry classification of a more or less 2D periodic image at a given signal to noise ratio.

While the translation symmetry of a Bravais lattice type is contained in a plane symmetry group, a Laue symmetry is just a point symmetry that includes the symmetry of the Fourier transform itself.

## 5.2. Bayesian posterior model probabilities and confidence sets for crystallographic symmetry classifications

It is expected that the Bayesian posterior model probability update approach of equations (12) and (13) will be helpful for the recognition of genuine pseudo-symmetries in noisy 2D periodic images that exist either per design of the image or by the nature of the crystalline sample that has been imaged. This expectation is founded on the fact that equation (12) represents a product of normalized probabilities.

If we take, for example, the two images in Figure 1, it was demonstrated in section 3 that the parameters of a rectangular Bravais lattice can be readily extracted in reciprocal space even in the presence of Gaussian noise and a genuine translational pseudo-symmetry when one takes the amplitude map of the dFT into account, see Tables 2b and 3b. Both of these images will, therefore, obtain large Akaike weights for the rectangular Bravais lattice type by the application of equation (8). The Akaike weights for the square lattice type, on the other hand, will be very small for both images given the extracted lattice parameters in Tables 2b and 3b. This will settle the question of the prevailing translation symmetry in the presence of a genuine translational pseudo-symmetry.

As for the symmetry of the 2D periodic motif, the somewhat "squarish" appearance of the image in Figure 1b suggests that its Akaike weights (equation (8)) for plane symmetry groups $p4$ and $p4mm$ could be high, while being modest for plane symmetry group $pm$. We know, however, from the design history of this image that (i) plane symmetry group $pm$ and (ii) the rectangular Bravais lattice type are the correct crystallographic symmetry classifications for this image.

If we calculate the product of both types of Akaike weights for this image with equations (12) and (13), there should be a good chance that we obtain the correct crystallographic symmetry classification even in the presence of noise and a motif-based genuine pseudo-symmetry. This should be the outcome of the significant down weighting of the Akaike weights of plane symmetry groups $p4$ and $p4mm$ by the very small Akaike weight for a square Bravais lattice given the extracted lattice parameters of Tables 2b and 3b.

At least, this is the "promise" of the mathematical form of equation (12). To which extent this is indeed so in practical cases of interest needs, of course, to be demonstrated experimentally for more or less 2D periodic images with various genuine pseudo-symmetries and noise levels.

There are three different types of crystallographic symmetries in noisy 2D periodic images. Each of these symmetry types is hierarchic and there are several distinct hierarchy branches [6,21]. Accordingly, there are several different ways of summing Akaike weights up to confidence sets.

Creating confidence sets for each of the hierarchy branches and each of the symmetry types should be helpful for the recognition of genuine pseudo-symmetries. Again, it needs to be demonstrated experimentally for more or less 2D periodic images with various genuine pseudo-symmetries and noise levels to what extent this is indeed so in practical cases of interest.

## 5.3. Multi-model averaging for better predictions and safer conclusions

The common symmetry classification practice in crystallographic image processing [19,20] and electron crystallography [21] is currently characterized by attempts to infer the plane symmetry of a more or less 2D periodic image on the basis of the three traditional symmetry deviation quantifiers of electron crystallography (which are succinctly described in ref. [6]). That approach does, however, rely on judiciously set thresholds as already stated in the Introduction and Background section of this review and is, therefore, not objective.

As a result of the prevailing subjective practice, one ends up with one model description of the image only, which allows for only one set of lattice parameters and defines all pixel intensities within the asymmetric unit. (That unit is the part of the translation periodic motif from which all other pixel intensities within a unit cell are created by the application of the applicable 2D space group symmetries of the model.) The individual pixel intensities within the asymmetric unit are in effect the average over the whole image of all symmetry related pixel intensities according to the chosen crystallographic model description (i.e. plane symmetry group).

One cannot, however, be sure that one has selected the best possible model given the amount of generalized noise in the data as there is subjectivity in the model's selection when one chooses it in accord with the common practice.

The raffle ticket discussion of the previous section, where a set of model probabilities adds up to 100 %, allows one to take a broader view and feel comfortable with the spreading of a crystallographic symmetry classification over a range of models that form a set. This is because each of the models in the set possesses an objectively quantified amount of merit for representing the geometric-structural information in the noisy data optimally, while the model with the highest probability ($\Delta_i = 0$) is the one that does this job best in the sense that it minimizes the expected Kullback-Leibler information loss.

Since one ends up with several model descriptions in the information theory inspired approach of this review while following objective criteria, one also has several sets of unit cell parameters and pixel intensities for the asymmetric units, which one may call collectively individual model parameters.

It is a good idea to average over all of the related parameters (and predictions) from all of the models in a set on the basis of their respective Akaike weights, $w_i$, as defined in equation (8). The main advantage of multi-model averaging is summed up by Akaike himself in his statement: *"If the choice of one single model is not the sole purpose of the analysis of the data the average of the models with respect to the approximate posterior probability C exp {(½) AIC(k)} will provide a better estimate of the true distribution of Y."* [106], where C stands for a constant, *AIC(k)* stands for the $\Delta_i$ values of a set of models, and Y stand for a set of observations.

In cases of geometric AICs and more or less 2D periodic images, the observations are the unit cell parameters and the gray levels of the individual pixels. There is typically, as already mentioned above, only one image (or at most a few images) with a given generalized noise level to be classified.

Multi-model averaging safeguards against the obtaining of results that may actually refer to symmetries that would not survive extrapolations to vanishing feature extraction uncertainties (or a zero generalized noise level, in other words). Such slightly broken symmetries may already be present in a sample when it is, for example, a mixed crystal where various transition metal atoms substitute for each other at sites of otherwise moderately high symmetry as it happens in many lower symmetric minerals. It may be difficult (or with currently available technologies quite impossible) to distinguish between such substituted transition metal atoms in imaging experiments reliably so that the above mentioned multi-model averaging safeguard could be important in order to avoid conclusions that are wrong. Multi-model averaging will typically result in few to zero symmetries but provides the *statistically best* value of a parameter in the presence of generalized noise.

To illustrate averaging over model predictions and estimated model parameter values in the context of this review briefly, let us assume that the Akaike weights for oblique, rectangular (primitive), rectangular-centered, and square Bravais lattice types are 10 %, 50 %, 25 %, and 15 % for a noisy 2D image with many repeats of a more or less translation periodic motif. The averaged unit cell angle is then for example obtained by multiplying the individual "angle parameter estimates" of the four models by weights 0.1, 0.5, 0.25, and 0.15, respectively, and the subsequent summing of the four products.

The last of these products is simply 90° times 0.15 for the square Bravais lattice type. Note that the angle between the two unit vectors of equal length of the primitive subunit of the corresponding (two times larger) rectangular-centered unit cell [6,98] needs to be used in the averaging procedure in the penultimate one of the four products. This angle will be somewhat close to 90° for our example as the corresponding Akaike weight for that particular translation symmetry model is 25 %.

While the unit cell angle of the oblique Bravais lattice is also somewhat close to 90°, it is precisely 90° for the rectangular (primitive) Bravais lattice type. The averaged angle will consequently be rather close to 90° since the exact 90° value has a combined probability of 65 % for our example. Note that while the rectangular (primitive) Bravais lattice type is for our example the translation symmetry model that minimizes the expected Kullback-Leibler information loss (since it possesses with 50 % the largest Akaike weight), the averaged unit cell angle is not restricted to be exactly 90°.

### 5.4. Acknowledging model selection uncertainties in qualitative ways

Akaike weights [12-15] (equations 8,12,13) are also useful for qualitative acknowledgments of model selection uncertainties. Obviously, if the best model has only a probability of 30 % and the second best comes in at 28 %, the selection of the first model is quite uncertain given the fact that geometric AICs are first order approximations for small Gaussian noise levels. One should, therefore, not rule out the second best model and, at least, communicate that model's probability (and even better all of the probabilities of the models in the set) for inclusion in databases.

When better experimental data (with improved signal to noise ratio) and more accurate processing algorithms become available later on in the future, these two models may either change their respective likelihood rankings or the better model may command a higher percentage of the probability of being the Kullback-Leibler best model. For the time being, one is however stuck with two models that possess a relative likelihood ratio of nearly unity so that the evidence for the $\Delta_i = 0$ model is not much stronger than that for the $\Delta_i > 0$.

To illustrate this with the so far only available examples from the literature, we turn to ref. [4]. Liu and coworkers give in the earlier version of their 2004 paper [4b] details of frieze group classifications for both a walking humanoid avatar and a walking human being. With an inconsistency that is probably due to ignoring the applicable crystallographic origin conventions [31] and further discussed in appendix D, the walking avatar features most likely frieze symmetry $\not{p}2mg$ with a comparatively small model selection uncertainty.

The time series data for the walking human being is much noisier so that the least-squares residuals for the disjoint frieze symmetries $\not{p}2mg$ and $\not{p}2mm$ [109] are nearly equal to each other. A large model selection uncertainty results, therefore. The reason for this could well be the simultaneous existence of genuine pseudo-symmetries in the form a glide-line and a horizontal mirror-line in the recorded time series data, see appendix D for further discussions.

Returning to the two images of Figure 1, the model selection uncertainty for the noise-free image to the left (Fig. 1a) is clearly going to be much smaller than for the image where Gaussian noise has been added (Fig. 1b.) Note finally that the question of model appropriateness is irrelevant in the context of this review because the mathematical frameworks of 2D and 1D crystallography [21,31] are without any doubt appropriate for crystallographic symmetry classification of more or less periodic crystal patterns.

# 6. Summary and Conclusions

Geometric Akaike Information Criteria and associated Akaike weights for generalized noise-level dependent crystallographic symmetry classification of 2D images that are more or less periodic in 2D (or 1D) and considered to constitute 2D data planes have been reviewed. These kinds of classifications are always fuzzy and, in a sense, preliminary since images with reduced generalized noise levels may become available in the future. In other words, these kinds of classifications are never *definitive* and static in all real world applications in compliance with Kanatani's dictum.

While this review concentrates on more or less periodic crystal patterns in two dimensions (and mentioned such patterns in one dimension only briefly on a few occasions), it goes without explicitly saying that the outlined approach is in principle also applicable to crystal patterns of dimensions three to six.

It was demonstrated by an example that pseudo-symmetries present challenges to extraction algorithms for geometric-structural features from more or less 2D periodic images as well as to their subsequent crystallographic symmetry classifications. Pseudo-symmetries in 3D and the problems they cause in mainstream single crystal X-ray crystallography are briefly discussed in appendix C. It is noted in that appendix that there are so far no statistical descriptors in mainstream 3D crystallography beyond the Hamilton test, which is a form of null hypothesis testing, that are dedicated to dealing with Kanatani's comments. Similarly, there is so far no systematic procedure to deal with genuine pseudo-symmetries in 3D on the basis of noisy diffraction data.

The point is also made repeatedly in appendix C that crystallographically mis-classified 3D crystal structures could essentially no longer be found within crystallographic databases as soon as the objective information theory based approach of this review was implemented and symmetry classifications were allowed to spread over several classes as a function of the generalized noise level of the experimental data. Such a spreading would allow for an objective reporting of the results of crystal structure determinations, but may not be necessary for very highly symmetric and very well characterized atomic arrangements, where there is hardly any lingering doubt about the validity of the reported structure. For lower symmetric and poorly characterized atomic arrangements (as in many biopolymers), on the other hand, the spreading over several crystallographic symmetry classes would be helpful to the users of the databases as uncertainties about the structures' validity are faithfully/objectively reported. When objectively better crystal structure determinations become available in the future (at lower generalized noise levels), the spreading would allow for a simple updating of the database entry rather than a re-classification.

Crystallographic model selection uncertainties were illustrated in a qualitative manner on the basis of results from the single relevant experimental study (in 1D) in the literature that the author of this review is aware of after quite substantial background searches. Multi-model inferences and averaging were also discussed.

The combining of Akaike weights for Bravais lattice types, Laue classes, and plane symmetry groups should enable successful crystallographic symmetry classifications even in the presence of manifest pseudo-symmetries that exist per design of an image or pre-exist within a crystalline sample that has been imaged.

Despite the lack of a guarantee that Kanatani's geometric AIC approach will work well for fuzzy but quantitative crystallographic symmetry classifications, the members of the applied crystallography and computational symmetry communities are hereby invited to test them out on the basis of the above listed equations. Demonstrated success in that endeavor could lead over time to a widespread adaptation of the information theoretic approach (as briefly outlined in this review for the 2D case) to mainstream 3D periodic crystallography and its higher dimensional extensions.

Appendix A is directed at the applied crystallography community and briefly assesses the potential of the main proposal of this review in connection with the extraction of grain boundary structures from atomic resolution images that are more or less periodic in 1D. Appendix B distinguishes between pseudo-symmetries of different types. (Experimental noise turns genuine symmetries in data that was collected from a crystal structure into pseudo-symmetries of the second kind, which can be very hard to distinguish from genuine pseudo-symmetries.)

Appendix D provides comments of a crystallographic nature on the only relevant experimental study in 1D from the literature that has employed a geometric AIC. These comments are peripheral to the main topic of this review but still worthwhile making for the benefit of the computational symmetry community.

## Acknowledgments

Professor Bryant York of the Computer Science Department of Portland State University (PSU), Andrew Dempsey and Paul DeStefano of PSU's Nano-Crystallography Group are thanked for critical proof-readings of the manuscript. PhD student Paul DeStefano is also thanked for the illustrative graphic that accompanies the abstract text in the Table of Contents of the corresponding special issue on "Mathematical Crystallography" of *Symmetry*. Professor Werner Kaminsky of the Department of Chemistry of the University of Washington at Seattle is thanked for his critical proof-reading of appendix C.

Emeritus professors Kenichi Kanatani (of the University of Okayama/Japan) and Wolfgang Neumann (of Humboldt University Berlin/Germany and 2017 Carl Hermann Medal Laureate) are thanked for helpful discussions and encouragements to pursue the lines of research that led to this review, which PSU's faculty enhancement program supported financially. The publication of this article in an open access journal was funded by the Portland State University Library's Open Access Fund.

## Conflicts of Interests

The author declares no conflicts of interests of any kind. The funding sponsors had no role whatsoever in the design of this review; in the collection, analysis, and interpretation of the data; in the writing of the paper, and the author's decision to publish it in the above mentioned special issue of *Symmetry*.

## Appendix A: Opportunities and background in the context of 1D periodic 2D images as obtained from plane edge-on projected 3D grain boundaries

Reference [36] demonstrates the application of a computational tool for the automatic extraction of the five macroscopic grain boundary parameters (geometric degrees of freedom) [43] from atomic resolution STEM images of GaAs and other materials with zinc-blende structure. Just like the proposed tool of ref. [28] for automated symmetry classifications of essentially defect-free 2D periodic crystals on the basis of their atomic resolution microscope images, this very recent tool could be enhanced by the incorporation of core ideas of this review as applied to 1D periodic cases.

Note that a real grain boundary possesses four microscopic [43] parameters that influence its physical properties in addition to the five macroscopic parameters that the new tool is able to extract automatically. The four microscopic grain boundary parameters are somewhat restricted by the CSL index that results from the macroscopic grain boundary parameters [43].

Note also that grain boundaries with the same tilt angle, tilt axis, and crystallographic interface plane (i.e. the same five macroscopic parameters) can feature different types of frieze symmetries in edge-on projections, which correspond to different low-energy structures around the interface at the atomic level [37]. In *bi*-crystallographic terms, different types of frieze symmetries result for the same grain boundary in edge-on projections from the sectioning of a 2D periodic dichromatic pattern or complex with the same CSL index [37,44]. The sectioning process is also known as the scanning of a dichromatic space group of any dimension for different types of subperiodic groups [31,44]. (The 3D atomic arrangement around an interface plane results, for example, from the sectioning of a dichromatic space group into a layer group [31]. When a 2D periodic dichromatic pattern or complex is sectioned [44], a 1D periodic frieze symmetry group arises [37].)

Automated determination of the four microscopic grain boundary parameters (in addition to the five macroscopic parameters) along with the automated classifications of grain boundary segments [37] into frieze symmetries would be highly desirable in connection with trying to make progress within the above mentioned crystalline materials per design [16] paradigm. By means of the application of Pierre Curie's symmetry principle [110], one obtains different types of allowed physical properties, e.g. polar or non-polar, across grain boundary segments in dependence of their frieze symmetries. *Bi*-crystallography allows also for the derivation of symmetry dictated maxima and minima of the physical properties of the interface region [39].

Aberration corrected atomic resolution STEM imaging revealed, for example, that segments of the same $\Sigma$ 13 [001] (510) tilt boundary with different frieze symmetries in high purity $SrTiO_3$ accommodated significantly different amounts of dopants that substituted for titanium close to the interface [45]. These results are statistically significant as several tens of STEM images with the same frieze symmetry along the same grain boundary containing several hundreds of sectioned CSL unit cells were averaged in order to enhance the image contrast and obtain representative atomic arrangements for the differently sectioned CSL unit cells. For a related study on the same type of grain boundary with the same types of frieze symmetries in undoped high-purity $SrTiO_3$, the images of approximately 400 sectioned CSL unit cells in approximately 50 STEM images were averaged in direct space in order to reveal the characteristic atomic arrangements around the interface for each of the different frieze symmetry types [33]. Note that the orientation relationship and crystallographic interface plane, i.e. the five macroscopic parameters, of the grain boundaries in all of these related studies were highly precise because they were prepared with the high-temperature diffusion bonding technique [33,37,45]).

Small geometrical deviations from the geometry of sectioned CSL lattices are in general grain boundaries with plane interfaces thought to be mediated by interfacial quasicrystallinity [41] so that they are in principle amenable to descriptions with the concept of periodic (and symmetric) "quasicrystal approximants".

The term quasicrystal approximant, as coined by Christopher L. Henley in the late 1980s, refers to an ordinary 3D periodic crystal with a very large unit cell and an atomic structure that is "to some extent" indistinguishable from that of a genuine quasicrystal. The atomic arrangement of a fragment of the approximant's unit cell occurs also in a quasicrystal, which is translation periodic in six dimensions.

The real numbers and ratios that characterize the ordinary unit cell geometry of approximants are "bracketing" the irrational numbers and ratios that are characteristics of the genuine quasicrystals which they are approximating. While the components of CSL transformation matrices are rational numbers for periodic grain boundaries, irrational components of such matrices are characteristics of quasi-periodic grain boundaries.

All general grain boundaries with plane interfaces can, thus, in principle be approximated to periodic high CSL index grain boundaries and *bi*-crystallography [31,37-41] is applicable to all of them. The predictions of frieze symmetry types in atomic resolution TEM images by means of 2D *bi*-crystallography [37] are in principle also valid for all high (and low) CSL index grain boundary approximants to general ("quasi-periodic" and "quasi-symmetric") grain boundaries with planar interfaces.

The CSL index for a sufficiently good approximation may, however, be very high so that the actual grain size might set practical limits to the applicability of the *bi*-crystallographic symmetry theory to general grain boundaries with planar interfaces. When finite temperature effects are included, high CSL index grain boundaries reduce effectively to low CSL index grain boundaries [42] as the five macroscopic grain boundary parameters become less well defined.

## Appendix B: Pseudo-symmetries

Pseudo-symmetry refers according to ref. [64] to "*a spatial arrangement that feigns a symmetry without fulfilling it*". Be aware of the distinction between *genuine* pseudo-symmetry, which is in accord to the definition of the IUCr at the URL given as ref. [46] and *non-genuine* pseudo-symmetry, which arises from the effects of random and systematic distortions on the symmetry operations that form space groups in 1D, 2D, and 3D. Such distortions occur unavoidably in any experiment and are considered to constitute generalized noise in Kanatani's sense [8-11] in a generalized imaging experiment. Genuine pseudo-symmetries are also referred to as pseudo-symmetries of the first kind. Non-genuine pseudo-symmetries, on the other hand, are referred to as pseudo-symmetries of the second kind [99].

For translational pseudo-symmetry in 3D, see refs. [48], [57] and [63]. For "pseudo-rotation/screw axis + pseudo-mirror/glide plane mediated" = motif-based pseudo-symmetry in 3D, see refs. [48] and [49]. For a discussion of pseudo-symmetries in the context of protein crystallography, see http://www.rcsb.org/pdb/help/viewers/jmol_symmetry_view.html.

Distinctions have also been made between local and global pseudo-symmetries [70]. Global pseudo-symmetry operators of the first kind are located at positions that allow for their approximate combination with existing pseudo-symmetries of the second kind so that the unit cell content acquires apparently a higher symmetric pseudo-space group. The results of such combinations are often complications in least-squares refinements of crystal structures form noisy 3D diffraction data and mis-classifications in crystallographic databases, as discussed below in appendix C.

Genuine global pseudo-symmetry has also been referred to as "Fedorov pseudosymmetry" (E. V. Chuprunov) and is reviewed with respect to relationships between atomic structures and physical properties in *Crystallography Reports* **2007**, *52*, 1–11. (There are 230 possible pseudo-space groups in 3D just as there are 230 genuine "Fedorov symmetry groups", which are outside of Russia referred to as "3D space groups".) The program PSEUDO at the website of the Bilbao Crystallographic Server (http://www.cryst.ehu.es/cryst/pseudosymmetry.html) facilitates the interactive elucidation of this type of pseudo-symmetry.

Genuine local pseudo-symmetry operators (of the first kind), on the other hand, are often referred to as being "non-crystallographic" and located at positions that do not allow for their approximate combination with pseudo-symmetries of the second kind so that higher symmetric pseudo-space groups cannot be formed. That kind of pseudo-symmetry has, therefore been referred to as "non-Fedorov pseudosymmetry" (G. M. Lombardo and F. Punzo, *J. Molecular Structure* **2014**, *1078*, 158–164).

Both global and local pseudo-symmetries may relate atoms of a molecule to atoms in another molecule in some approximate manner in the same asymmetric unit of a Z' > 1 crystal structure. The atoms do not necessarily need to be of the same kind as similar densities in an electron density map of a single crystal X-ray diffraction experiment may suffice. Oxygen atoms may, thus, be related to nitrogen atoms by pseudo-symmetry of the first kind if their respective coordinates are somewhat related by some approximate symmetry.

The term "non-crystallographic symmetry" is also used in the structural biology community, mainly (but not exclusively) in reference to local pseudo-symmetry of the first kind [73]. Depending on the particulars, the IUCr prefers to use the terms (genuine) pseudo-symmetry, local symmetry, and partial symmetry instead (*Online Dictionary of Crystallography of the IUCr:* http://reference.iucr.org/dictionary/Noncrystallographic_symmetry) whereby the definition for (genuine) pseudo-symmetry is as given above and at the URL listed as ref. [46]. Strictly speaking, non-crystallographic symmetry must be defined as a negation of crystallographic symmetry (Nespolo, M.; Souvignier, B.; Litvin, D.B. About the definition of "noncrystallographic symmetry", *Z. Kristallogr.* **2008**, *223*, 605–606) and not as a feigning of the latter. The usage of the term "non-crystallographic symmetry" in the structural biology community includes approximate symmetries that are strictly local and, therefore, not subject to the well known crystallographic symmetry restrictions [21].

An example for this is an approximate five-fold rotation point in a 2D periodic crystal pattern. Note that this rotation point supports an approximate symmetry only in its immediate surrounding as there are no five-fold site symmetries in any plane symmetry groups per definition [21]. That rotation point will, however, be present in each unit cell due the actions of the genuine symmetry operations of a plane symmetry group.

Non-crystallographic symmetry is very common in Z' > 1 structures and was estimated to be present in roughly one half of the protein structures that have been solved and refined at low resolution (Kleywegt, G.J.; Jones, T.A. Where freedom is given, liberties are taken. *Structure* **1995**, *3*, 535–51). Since protein molecules always crystallize with the inclusion of solvents, pronounced non-crystallographic symmetries may lead to an exacerbation of the bias of the $R_{free}$ value (Fabiola, F.; Korostelev, A.; Chapman, M.S. Bias in cross-validated free $R$ factors: mitigation of the effects of non-crystallographic symmetry, *Acta Cryst. D* **2006**, *62*, 227–238). This may contribute to suboptimal protein structure refinements from low resolution data.

If there is no genuine pseudo-symmetry and low noise data, the joint probability of "matching $w_{i,q}$ pairs" and "$w_{i,q,s}$ triples" of Akaike weights for Bravais lattice types and plane symmetry groups as well as for Laue classes will be high in equation (12) because each of the individual probability factors will be large. Such matching pairs and triples are defined by mutual crystallographic compatibility conditions.

The rectangular (primitive) Bravais lattice type and Laue class *2mm* are, for example, crystallographically compatible with each other and both are also compatible with plane symmetry groups *pm*, *pg*, *p2mm*, *p2mg*, and *p2gg*. (All of the symmetries that are components of a particular plane symmetry group are obviously compatible with each other.)

Genuine pseudo-symmetries in real world images represent, on the other hand, cases of non-matching pairs and triples of Akaike weights because they are only in some approximate manner compatible with the genuine symmetries of a corresponding hypothetical noise-free image.

## Appendix C: Pseudo-symmetry and other problems in mainstream 3D crystallography that have led to mis-classified entries in major crystallographic databases

### C1. Space group symmetry mis-classifications in major crystallographic databases

The crystallographic literature demonstrates clearly that genuine pseudo-symmetries and experimental noise (that turns genuine symmetries into pseudo-symmetries of the second kind) are complicating single-crystal X-ray crystallography structure analyses in 3D, but this appendix can only discuss some of the most influential papers in this very wide field.

Starting with William Laurence Bragg's original mis-assignment of the primitive cubic Bravais lattice type to sylvite (KCl), a historic review of incorrect structures that are associated with genuine pseudo-symmetries is given in ref. [59]. "Ambiguous" space group assignments account for about one half of the discussed cases in ref [59] and this appendix will be mainly concerned with analogous mis-classifications of crystallographic symmetries.

Mainstream 3D crystallography relies on experimentally obtained X-ray diffraction patterns consisting of Bragg peaks and associated background from a single crystal, the kinematic diffraction theory, Fourier transforms, and least-squares refinements. The crystals are supposed to consist of spherical atoms or point nuclei that undergo harmonic displacements with respect to their equilibrium position independently.

Compared to electron crystallography on the basis of HRTEM images that were recorded within the validity of the WPO approximation [103], this requires mainstream 3D crystallography to find a solution to the problem that the structure factor phase angles (phases for short) are not directly measurable but required for the solving and refining of a crystal's structure. The structure factor phases are in electron crystallography, on the other hand, obtained directly from the Fourier coefficient phase angles of the intensity of noisy 2D periodic HRTEM images [103] by means of crystallographic image processing [19,22].

As about 80 % of the "information" in a crystal structure resides in structure factor phases and only about 20 % in structure factor amplitudes [111], this is a distinct advantage of the direct space imaging approach. The tomography approach [22] combined with discrete goniometry [112-114] is also applicable in atomic resolution HRTEM in order to obtain information in 3D. Another peculiarity of the direct space imaging approach is that the magnitudes of the complex structure factors can be obtained directly from Fourier transforms of the images so that strictly linear least-squares refinements can be undertaken.

It goes without saying that crystals possess genuine symmetries only as spatial and temporal averages. When crystals are sufficiently large and perfect, the spatially and temporally averaged atomic arrangement within a unit cell can be idealized by a geometric-structural model that possesses a well defined space group symmetry. Deriving that space group symmetry on the basis of experimentally obtained data from a real crystal is, however, fraught with ambiguity because crystallographic symmetries are hierarchic and mathematical abstractions only. Due to noise in the extraction process of geometrical/structural parameters, genuine pseudo-symmetries may easily be mistaken for genuine symmetries (that turned into pseudo-symmetries of the second kind due to experimental noise and the particulars of utilized extraction algorithms).

It is highly instructive to discuss problems in mainstream 3D crystallography in the context of this review because more or less *subjective* classifications into Bravais lattice types, Laue classes, and space group types are made there as well on the basis of noisy experimental measurements as part of the standard procedures of the current state of affairs. The resulting crystallographic symmetry classifications and mis-classifications end up subsequently in crystallographic databases.

Based on a reasonable threshold, it was estimated in 2008 that some 6 % [48] of the entries in the open access world-wide Protein Data Bank (wwPDB often abbreviated as just PDB) [75] refer to structures with pseudo-symmetries. At least some, but possibly many of these pseudo-symmetric structures are mis-classified in the public biopolymer structure record and currently do not allow for the derivation of factual structure-function relationships. This is to a large extent due to the fact that *"structures in the PDB are based on a subjective interpretation of experimental data, which may itself be of variable quality, a process that can lead to errors with varying degrees of impact"* [76]. What is true for the PDB is also true for any other sufficiently large database [77-79] of mainstream 3D crystallography results.

Utilizing appropriately defined measures, the prevalence of genuine translational and inversion pseudo-symmetry in 211,162 crystal structure entries for organic and organometallic compounds in the Cambridge Structural Database (CSD) [77] was assessed by N. V. Somov and E. V. Chuprunov in the year 2009 [56]. Of the 60,707 entries with non-centrosymmetric space groups, approximately 19.8 % featured a pseudo-centrosymmetry. Approximately 4.7 % to 6.1 % of the analyzed structures featured a translational pseudo-symmetry. While the percentages of pseudo-centrosymmetry were higher for lower symmetric crystal systems (e.g. triclinic and monoclinic) than for their higher symmetric counterparts (e.g. tetragonal, hexagonal, and cubic), the opposite was true for translational pseudo-symmetry [56].

Up to 2 % of the single crystal X-ray crystallography structures of proteins in the wwPDB are suspected to fit potentially into higher symmetric space groups [48]. Many of these descriptions of crystal structures are not necessarily wrong. These crystal structures are often just reported in a space group that is a subgroup of some reasonably well fitting higher symmetric space group. In cases of molecule crystals, some of the 3D point symmetries that individual molecules possess may then not have been recognized during the X-ray crystallography analysis and remain unrevealed in the corresponding database records.

The review of ref. [49] reminds the reader also that (*i*) *"experimental measurements never establish the space group with absolute confidence. There are always physical uncertainties to be considered both in the positions and the intensities of the Bragg reflections"* and (*ii*) concludes that *"the R-factor statistics did not help at all to distinguish between the best symmetry and the underassigned symmetry"*.

Reference [50] states along similar lines and also while referring to the wwPDB that (*i*) *"the problems created by missed symmetry cannot be addressed using techniques based on quality control statistics such as $R_{free}$, the crossvalidation (CV) statistic introduced in 1992 on which so much reliance is placed today"* and (*ii*) issues a *"call to arms to the entire structural biology community so that the important, but entirely correctable problems"* which that paper discusses can be resolved as far as this is possible given Kanatani's dictum [8,9,99]. While the above mentioned CV index $R_{free}$ is described in detail in ref. [51], the review by P. G. Jones provides background on the most commonly used $R$ value and its weighted form ($R_w$) [82]. Hamilton introduced generalized weighted $R$ values ($R^G$ and $R''$) in order to facilitate null hypothesis tests concerning the question if the addition of refinement parameters enhances the validity of a structural model in a statistically significant manner [85a].

The generalized noise-level dependent crystallographic symmetry classifications that are proposed in the main body of this review could be adapted to 3D as part of the solution of these problems both during the experimental structure analysis procedures and at the database level. This is because there could essentially no longer be crystallographic symmetry mis-classifications within the major databases for the large class of structures that potentially fit into higher symmetric space groups once the objective information theoretical approach of this review is implemented in 3D. The small price to pay for this would be just the spreading of the entry of a small organic molecule, protein … intermetallic or mineral crystal structure over a range of crystallographic classes to which everybody would get used to over time.

Very well known crystal structures such as the so called structural prototypes of inorganic materials science and mineralogy, e.g. Cu, Mo, Mg, diamond, NaCl, CsCl, $BaTiO_3$ … should remain as they are classified right now, i.e. assigned to one crystallographic symmetry class only as there is no uncertainty to which class they truly belong. (Utilizing the Akaike weight concept, these structures have been classified with likelihoods exceedingly close to 100 % so that there is neither a need nor a basis to spread their entry over several crystallographic classes.) All of these structural prototypes are highly symmetric and a thorough review [52] revealed that inorganic materials with very high crystal symmetries are very rarely mis-classified.

If one deals with certain low-symmetry minerals, on the other hand, where different atoms substitute for each other and there are noticeable differences in the chemical composition depending on the place from which a mineral has been obtained with respect to another such place, a wide spread over crystallographic symmetry classifications and the inclusion of information on from where the mineral sample has come from into the database record would be in order.

As core of the crystallographic classification problems discussed above, ref. [53] identifies *the nature of human cognition, which is frequently influenced by preconceptions that may lead to fanciful results in the absence of proper validation"*, i.e. subjectivity, in other words. That subjectivity is bound to contaminate the structure validation process also as long as it is not done fully objectively, i.e. as long as the ideas outlined in the main body of this review have not been transferred to 3D and are not implemented in structure validation procedures.

Reference [53] states also that *"it would be useful if all data-processing programs took into account all possible supergroup/subgroup relations during the indexing and merging procedures and presented the suggestions to the users"*. There is obviously no need to restrict that idea just to the indexing and merging procedures of mainstream 3D crystallography.

One may as well solve and refine an atomic structure in all reasonable space groups within a particular symmetry hierarchy branch and quantify the relative likelihood of each crystallographic model by means of Akaike weights that are summed up into confidence sets, as discussed in the main body of this review. In cases of strong genuine pseudo-symmetries [46], one could also include quantifications of the relative likelihood of models that are not within the same symmetry hierarchy branch and part of a different confidence set.

After these comments, it is instructive to take up the review of papers that report mis-classification in the field of mainstream 3D crystallography again. This is done mainly to provide material for more comments that connect this appendix to the main body of this review.

There are many structures of small molecules in the CSD [77] which were refined in apparently "wrong" space groups [52,54,55,57-59,74]. Approximately 3 % of the entries in this database were estimated in the middle/late 1980s of the last century to feature unnecessarily low symmetries [54,55]. These problems are often associated with difficulties in reliably distinguishing genuine symmetries from genuine pseudo-symmetries when experimental noise levels are moderate to high.

The percentage of "problematic" structures that were published in the Journal *Inorganica Chimica Acta* (and entered into the CSD) was in 2005 approximately 3 % [57]. Approximately 16 % of these 260 structures in that journal had already been corrected before 2005 and ref. [57] attempts to do the same job for another 20 % of these structures, so that 167 "dubious" structural records that originated from that journal remained at that time in the CSD.

It was estimated that 17.7 % of the 260 problematic structures in that journal featured *"non-space group translations"* [57]. Such translations do not represent the translations that are really present in a crystal and are synonymous with translational pseudo-symmetry [46]. The real translations in these problematic structures are typically associated with weak reflections and/or systematic absences (extinctions) so that they are easily overlooked by a crystallographer who works with very noisy experiment data.

Note that this is more or less analogous to the "overlooking" of weak Fourier coefficient amplitudes by program C in its default setting as discussed in section 3 of the main body of this review. The relative weakness of the (1,0) and (2,0) amplitudes with respect to the (3,0) amplitude in the discrete Fourier transform of the images in Figs. 1a and 1b is caused by the translational pseudo-symmetry that was designed into the noise-free version of the image in Figure 1a.

Reference [58] lamented in 1995 that many journals which publish mainstream 3D crystallography results *"are relaxing their standards in many ways"*. This includes *"relegating crystallographic results to footnotes or even to supplementary material, selecting referees with little or no experience in crystallography and making it quite clear to their readers that, basically, any crystallographic details beyond a drawing of the molecule are unnecessary"* [58].

Richard E. Marsh has, therefore, *"no doubt that the percentage of incorrect results"* in those journals *"is appreciably larger than 3 %"* [58]. This criticism does, of course, not apply to the journals of the International Union of Crystallography (IUCr).

The analysis of 17,503 protein structures in the wwPDP that were obtained by means of single crystal X-ray crystallography concluded that the prestige (and alleged impact) of the journal in which these structures were published did not correlate positively with the quality of the structures as determined by the combination of nine complementing metrics [60a]. Due to a large percentage of protein structures with quantified quality below the overall average that they had published, the journals *Cell*, *Science*, *Molecular Cell* and *Nature* were ranked at the bottom of 30 journals in which the protein structure determination results had appeared. Note that this does not imply that the scientific conclusions on the basis of the published protein structures are invalid. What can be said on the basis of this ranking is that these structures are of restricted utility to the wider scientific community. Among the possible causes for this somewhat surprising result, the authors of that study point to *subjectivity* of referees and limited time and resources that some journals dedicate to the reviewing process [60a]. In the words of the authors of that study: *"the rush to publish high-impact work"* helps to explain *"the proliferation of poor-quality structures"* [60b].

Crystallographic structure validations have much improved with the implementation of significantly higher standards for publications in the journals of the IUCr and mandatory checks of deposited Crystallographic Information Files (CIFs) with sophisticated software for possible inconsistencies prior to publication and uploading to crystallographic databases. (The journal *Acta Crystallographica D* of the IUCr appeared in the top third of the above mentioned ranking of 30 journals that published protein structures as derived by single crystal X-ray crystallography [60].)

Anthony Spek describes great computer programs for such structure validations in refs. [65,66]. One of these two papers also reminds the reader of the importance of employing the crystallographic origin conventions (see also appendix D in this connection). That paper states that unrecognized genuine *"pseudo-symmetry can give rise to structures which initially appear to be plausible, but which have atoms or molecules misplaced with respect to the true symmetry"* [65].

The *"holy grail of structure validation"* would, according to Spek, be based on software tools that utilize *"objective criteria"* [66]. When the information theory approach and Kanatani's geometric AICs (as adapted to 2D crystallography classifications in the main body of this review) are eventually incorporated into future software tools for objective crystallographic structure validations in 3D, one might as well make provisions for databases to accept validated structures with noise-level dependent ("fuzziness-quantified") symmetry classifications.

Reference [66] also states that for only 384 out of 35,760 small molecule structures that were submitted to the CSD between 2006 and early 2007, the software on the submission sites for the journals of the IUCr indicated that a space group change was recommended. So the good news implied by ref. [66] is that from the year 2009 onwards, mis-classifications in this small molecule database might be below 1 %.

Note that (*i*) the subscription based CSD [77] and the open access Crystallography Open Database (COD) [78,79] possesses many more entries than the wwPDB and (*ii*) that corrections of the space groups, Laue classes, and Bravais lattice types of crystal structures often result in significant changes in both bond lengths and angles. Each of these corrections constitutes a crystallographic symmetry re-classification.

The small molecules in the CSD and COD are typically over-determined to a large extent in a single crystal X-ray diffraction experiment because there are many more Bragg reflection intensity measurements than there are free parameters of the atomic structures.

In single crystal X-ray protein crystallography, atomic resolution [80] is, on the other hand, often not achieved because the crystals do not diffract to below 0.12 nm. A plot of the number of observed reflections per atom for all X-ray crystal structure models in the wwPDB that were obtained from studies with a resolution < 2.5 Å peaked in 2011 at approximately seven [115]. For studies with a resolution $\geq$ 2.5 Å, the corresponding number at the peak of the distribution went down to approximately three. Since the number of experimental observations falls with the cube of the resolution, the crystal structure determination problem ceases to be over determined at low resolution.

One then uses prior knowledge in the form of restraints during the refinement of the model's geometric and thermal vibration properties. Such prior knowledge comes often from similar structure models in the wwPDB. In a typical refinement of a protein structure in a low resolution study, there may be many more restraints than there are actual physical observations in the form of quantified Bragg peak intensities [115].

Proteins are also much larger than small organic molecules so that larger $R$ values [116] result. For protein crystals, good $R_1$ values may be in the 15 % to 30 % range depending on the resolution of the data and the amount of solvent that remained within the crystals. Small molecule crystal structures with approximately 200 independent atoms, on the other hand, should be refined to $R_1 \leq 7$ % with an "allowance" for disorder of an extra 0.5 % [116].

The crystallographic phase problem can for protein structures not be resolved with direct methods so that Carl-Ivar Brändén and T. Alwyn Jones felt compelled to choose *"Between objectivity and subjectivity"* as title of their 1990 *Nature* paper [81] on protein crystallography. The records in the subscription based CSD and the open access COD do not contain biopolymers and should, therefore, be much less often mis-classified in the crystallographic symmetry sense than those in the open access wwPDB.

The curators of the wwPDB are well aware of protein structure specific problems and have taken multiple steps to address them. They have done comprehensive re-evaluations of their entries from 2008 onwards, provide now Internet based software systems for structure validations, and updated the file format of their structure entries to PDBx (which is based on the mmCIF format of the IUCr) [76].

In crystals with more than one molecule (formula unit) per asymmetric unit, pseudo-symmetry is rather widespread [67,68] and those crystals constituted between 8.8 % [69] and about 11 % [70,71] of all structures in the CSD in the year 2006. A genuine pseudo-centrosymmetry may in space group $P1$ easily be mistaken for a genuine inversion center [62]. The root-mean-square [82] deviations of the two chiral molecules in a $Z' = 2$ structures from their hypothetical counterparts in space group $P\bar{1}$ may be as low as 0.07 Å [72]. Jones provides in ref. [82] a re-classification of an inorganic (triclinic) $Z' = 2$ structure with space group $P1$ into its (monoclinic) minimal translationengleiche supergroup [117] $Cc$.

## C2. Reasons why $R$ values and similar "pure distance measures" are not helpful in crystallographic symmetry classifications

Reference [74] refers to non-biopolymer structures (so called small molecules) and distinguishes between *"quality structures"*, *"fuzzy structures"*, *"incorrect structures"*, and *"junk structures"*. While quality structures do not need to be discussed in this appendix, junk structures are what their name implies and could only be refined to *"R values well above 0.15"* [74].

Fuzzy structures *"are firstly and primarily characterized by good R values."* [74]. Incorrect structures *"have all of the same characteristics of the fuzzy group, including low R values, and so the two are often hard to distinguish"* [74]. Further characteristics of fuzzy structures are that atoms *"have been constrained or restrained in some fashion"* during the least-squares refinement process and possess unreasonable thermal vibration parameters [74].

The $R$ values in the quotes above (and in this review in general) refer to very popular measures for the "disagreement" between a least-squares refined atomistic model of the content of a unit cell of a crystal and the observations from that crystal in an X-ray diffraction experiment [118]. This "disagreement index", "residue", "$R$ factor", or "residual" is the normalized sum of the absolute value of the differences between the calculated structure factor magnitudes and the observed structure factor magnitudes

$$R = \frac{\sum \left| |F_{obs}| - |F_{cal}| \right|}{\sum |F_{obs}|} \quad (14),$$

whereby the latter are obtained by taking the square root of the intensity of the observed diffraction spots and the subscripts $_{obs}$

and $_{cal}$ stand for "*obs*erved" and "*cal*culated", respectively. (Note that using the square root of the reflection intensity introduces non-linearity into otherwise linear least-squares refinements and Hamilton's test.) The normalized sum of equation (14) is often multiplied by 100 % and the $R$ value that is defined by this equation is also referred to as $R_1$.

In the context of Kanatani's comments on symmetry as continuous and hierarchic features [3] (as briefly discussed in section 2 of the main part of this review), $R$ values are "pure distance measures" and, therefore, of limited use for deciding which atomistic symmetry model is best within a set of models that are within the same symmetry hierarchy branch.

Being in the same symmetry hierarchy branch means that the atomistic models are related to each other by subgroup and supergroup relationships. As far as 3D crystallography is concerned, the definitive reference text on such relationships is a publication of the IUCr [117]. Reference [119] applies these relationships to systematic crystal chemistry in the form of Bärnighausen trees.

With necessity, exclusive crystallographic symmetry classifications have to be *subjective* when based on $R$ values (and similar pure distance or disagreement measures) *alone*. It should, therefore, not have come as a surprise in the previous section that good $R$ values do not safeguard against crystallographic symmetry mis-classifications.

As a matter of fact, it is well known that reduced residuals ($R$ values) can be obtained from the fitting of models with more parameters to experimental data that contain negligible systematic errors and approximately Gaussian distributed random errors. An illustration for this is the structure model fitting in a crystallographic subgroup of a reasonably well fitting supergroup. Reference [120] reports, for example, for the mineral thortveitite, $\{Sc(Y,Fe)\}_2Si_2O_7$, weighted $R$ values (on the basis of the normalized differences of observed and calculated sums of structure factor squares) of 3.25 %, 2.83 %, and 2.79 % for space groups $C2/m$, $Cm$, and $C2$, respectively.

In Hamilton's own words: *"the model with the fewer restraints, that is, with the greater number of parameters, can usually be made to fit the data better than can the more restraint model"* and *"the model with the greater number of parameters can always be made to fit the data at least as well as the model with the fewer parameters, provided that the parameters in the latter are a subset of those in the former"* [85a]. (Note in passing that he italicized "always" for emphasis in the second quote.) When a structural model is incorrect, the $R$ value for a maximal subgroup of a space group, e.g. $R_{I4} = 19.5$ %, can be higher than its counterpart, e.g. $R_{I422} = 18.9$ %, for that space group as demonstrated by two refinements of a protein from the same low-resolution single crystal X-ray diffraction data [121].

Note that $R$ values are "somewhat related" to the first term in any Akaike Information Criterion [105,106] (including geometric AICs [8-11]) as discussed in sections 4.1 and 5.1 of the main body of this review. These first terms are always "model accuracy/disagreement measures" or "pure distance measures" depending on one's viewpoint, but all first order AICs have also a second term that corrects for biases, takes the complexity of the model into account, and provides a punishment for fits with too many free parameters.

The other key feature of all AIC applications is the "entertaining" of multiple models that individually possess quantified probabilities for representing experimental data with a certain noise level. It is, therefore, not only the goodness of the fit of a model to the data that counts but also the complexity of that model (and the prevailing signal to noise level).

Model selections based on AICs are *objective* because these criteria are based on rigorous mathematics and very deep foundations (which are considered to be beyond mathematical proof) such as the expected Kullback-Leibler information loss when a model is used to represent reality. The two terms of AICs make them parsimonious implementations of Occam's razor that can straightforwardly be calculated in order to "escape" from the *subjectivity* trap.

When systematic errors are negligible with respect to random errors, estimated standard deviations on bond lengths and angles are useful measures of the precision with which a crystal structure has been derived [82]. Because high parameter estimate precisions do not guarantee high parameter estimate accuracies, structures that were refined in the wrong space group may harbor significant unrecognized systematic errors while featuring reasonable $R_1$, $wR_1$, $wR_2$, and goodness of fit on $F^2$ values [122].

As far as the publications of the IUCr are concerned, references should be made to the standard uncertainty (s.u.) of a derived crystallographic quantity rather than to its estimated standard deviation (e.s.d.) [83,84]. These statistical measures are, however, also of limited use for deciding if a crystal structure is better described in a higher or a lower symmetric space groups when these groups are within subgroup-supergroup relationships.

As a matter of fact, there are so far no statistical descriptors in mainstream 3D crystallography beyond Hamilton's null hypothesis test [85a] that are dedicated to addressing Kanatani's comments [3]. Similarly, there is so far no systematic procedure to effectively deal with genuine pseudo-symmetries with respect to crystallographic symmetry classifications in 3D on the basis of noisy diffraction data.

## C3. Suspected pseudo-symmetry mediated space group symmetry mis-classification of a highly topical material

Metal-organic framework (MOF) compound NU-1000 (catena-[(µ8-4,4',4'',4'''-Pyrene-1,3,6,8-tetrayltetrabenzoato)-bis(µ3-oxo)-bis(µ2-oxo)-tri-zirconium], CSD entry number 955328, COD entry number 4120127) [86] has probably been mis-classified (after encountering severe problems in the refinement [87]). According to Richard L. Harlow, the published structure of this MOF is *"fuzzy"* at best [74] since a total of 133 restraints were placed on the positional and vibrational parameters of the identified atoms [86]. (As mentioned above, restraints are expressions of *presumed* prior knowledge that are treated in a least-squares refinement as if they were experimental observations.) Dore Augusto Clemente would probably classify this structure as both being characterized by *"non-space group translations"* [57] and incorrect.

Because Bränden and Jones *"strongly object to publication of structural work ... in the form of a cartoon"* [81], they would probably not be happy with the way the structure of this MOF is depicted/described in many papers that allege to report the key features of the crystal structure of NU-1000 according to its current CSD and COD entries. See refs. [88-91,95] for a small selection of such papers. The structure of this MOF is in the form of a cartoon allegedly characterized by *"exceptionally wide (31 Å) mesoporous channels extending throughout the structure"* [88].

The $R_1$ value of the single crystal X-ray crystallography study (on the basis of all hexagonally indexed reflections) was reported to be 13.17 % [86], but was so low only because the SQUEEZE function [92] of the well known OLEX2 [93] program had been utilized to remove a significant amount of experimentally observed electron density from the meso-scopic channels in NU-1000 during the original solving and refining of this structure [86].

Note that *Acta Crystallographica* suggested $R_1 \leq 7$ % as criterion for a reasonably well refined small molecule crystal structure in the year 2000 [116]. Participants of a conference on the *"Critical Evaluation of Chemical and Physical Structure Information"* considered non-biopolymer crystal structures with $R_1 > 10$ % as *"suspect"* in the year 1974 [59]. The goodness of fit on the square of the structure factor amplitudes was for NU-1000 as high as 1.737 [86], while is should ideally have been close to unity [122].

It is, therefore, somewhat doubtful if these two quantitative measures for the alleged "correctness" of the published structure of NU-1000 and the removal of observed electron density by electronic means can lead to *"that 'warm happy feeling' of confidence in the validity of the scientific work and the results presented"* that the participants of the above mentioned conference were talking about [61,123].

In other words, in order to obtain somewhat acceptable refinement results, partially long-range ordered/disordered material [86] needed to be "squeezed off" the meso-scopic channels so that they became apparently *"exceptionally wide"* [88] (and completely *empty* per definition of the word "channel"). The fact that these channels have been depicted to be empty in the form of cartoons in refs. [88-91,95] seems to be a direct consequence of this particular step in the original single crystals X-ray crystallography structure determination of this MOF [86].

Because these meso-scopic channels are in reality *not* completely empty (or are not *"exceptionally wide"* in other words) and the structure of NU-1000 is probably quite different from what is described in ref. [86], it is conceivable that some of the conclusions in refs. [88-91] are somewhat questionable (in spite of having been co-authored by one of the 2016 Nobel Prize winners in Chemistry). This does by no means imply that these conclusions are necessarily wrong. What can be said about them is, however, that one cannot report valid relationships between the atomistic structure of NU-1000 and its chemical/physical properties when there are serious doubts about the former.

The author of this review was given a few low-dose Z-contrast STEM images of NU-1000 for crystallographic analysis that show evidence for the co-existence of three different domains of low space group symmetry that are related to each other by a three-fold pseudo-rotation of approximately 120° about [001] and project along this axis to plane symmetry group *p2gg* or one of its translationengleiche subgroups [94a]. Much of the very high hexagonal symmetry that this MOF allegedly possesses according to ref. [86] (space group *P6/mmm*) is, thus, probably a consequence of pseudo-symmetries.

Note that this explanation is in agreement with Spek's assertion that *"pseudo-symmetry ... may result in partially*

*disordered structures when described with respect to the pseudo-symmetry element"* [65]. Also note that the probabilistic (fuzzy and generalized noise-level dependent) classification into a range of plane symmetry groups, e.g., *p2gg* and its translationengleiche subgroups, that this review proposes would obviously be very helpful for a quantitative communication of the preliminary analysis of the STEM images from this MOF.

One of the STEM images that the author of this paper was given has recently been published in ref. [95]. The faintly visible weak extra spots in the discrete Fourier transform of that image, which cannot be indexed on the basis of the alleged hexagonal lattice [86], have neither been mentioned nor discussed there [95]. (The STEM image itself can be accessed openly and directly at the URL [95] that is provided in the list of references. An indexed version of this image can be assessed at the URL given in ref. [94b].)

Whenever domains of penetration twins or "drillings" in crystals are very small, their "signature telltales" in X-ray diffraction patterns are very broad peaks [64]. When these peaks possess in addition low intensities, they may be easily overlooked in such diffraction patterns because simultaneously broad and weak peaks may be "buried" in experimental noise.

The wavelength of the electrons in the STEM study of NU-1000 being more than 60 times shorter than the wavelength of the single-crystal X-ray crystallography study [86] may have made the difference in resolving weak extra reflections that are nearly impossible to detect with Cu Kα radiation.

This author's crystallographic image processing [19,20] analysis also showed that there is evidence for positionally ordered material in the meso-scopic channels of this MOF, which are depicted as empty in the corresponding structure cartoons [88-91,95]. To make matters worse, virtually everybody else seems to assume that the meso-scopic channels are indeed completely empty and there have been numerous attempts to incorporate small molecules (including rotaxanes [89], catenanes [90], and fullerene derivates [91]) into these channels.

The alleged $C_{88}H_{44}O_{32}Zr_6$ asymmetric unit of this MOF is probably also underreported as far as its chemical composition is concerned. This is because more than one half of the experimentally observed electron density per unit cell [94a] has been removed from the single crystal X-ray crystallography analysis with the SQUEEZE function [92] of the OLEX2 software [93] (as already mentioned above).

This removal of a substantial amount of material from the analysis by electronic means has been faithfully reported in a qualitative way in the supporting material to ref. [86] and is quantifiable from information in the CIF that accomplished the structure determination of NU-1000 there. The removal of that material means, of course, that the systematic name of this compound is probably also in need of a revision.

The crystallographer behind the single crystal X-ray crystallography analysis of NU-1000 [86,87] was so kind and diligent as to leave comprehensive comments in the CIF and made structure factor amplitudes with hexagonal lattice indexing available as part of the openly accessible CIF in the supporting material to that paper [86] so that it was straightforward for Werner Kaminsky to reanalyze this crystal structure [94a].

The comments in that CIF make it clear that there is strong evidence for the meso-channels being partly filled by long range ordered material of an unknown chemical composition. Complementing pieces of information to this effect are contained in several comments within the CIF that is part of the supporting material of ref. [86].

Most non-crystallographers are, however, never going to read an individual CIF and are likely to believe instead that the cartoons which have been published multiple times in peer reviewed journals, see e.g. [88-91,95] for a very small selection, are complete and faithful representations of the results of the single crystal X-ray crystallography structure analysis that only appeared in the supporting material of ref. [86].

As it is typical for the present time, the original recordings of the X-ray area detector (so called X-ray diffraction images) are not available as part of the public structure record of this MOF so that new single crystal diffraction experiments are needed to reveal the full structure of NU-1000. Due to the rather large lattice constants and suspected small domain sizes of this MOF, it might be best if these experiments were to utilize synchrotron radiation.

In the meantime, comments should be added to the structural record of NU-1000 in the major databases that there is (*i*) probably a combination of a motif-based (i.e. "six-fold rotation plus mirror planes") pseudo-symmetry with a translational pseudo-symmetry in the analyzed crystal of ref. [94a] and (*ii*) generally ignored evidence for long-range ordered material in the allegedly *"exceptionally wide"* [88] channels.

The possible co-existence of three domains in the particular crystal that served as sample for the single-crystal X-ray crystallography study of ref. [86] may either be typical for this material or an unlucky coincidence as other crystals with a single domain throughout a sample may exist.

It was also theorized recently that the original synthesis procedures of NU-1000 [86,88] may lead to the formation of "heterogeneous crystals" where NU-1000 and "NU-901-like" structures co-exist [96] in the same sample and some of the exceptionally wide channels are filled with Zr containing nodes in positions and orientations that break the alleged hexagonal symmetry. The proposed heterogeneity of NU-1000 "pseudo-crystals" that were synthesized according to the procedures described in refs. [86,88] could possibly be an explanation for unusually strong local variations of the observed lattice parameter in low electron dose STEM images of NU-1000 [94] that seem to comply with a paracrystal model [97]. Apparently "phase pure" crystals of NU-1000 have been obtained by a new synthetic route [98] recently so that new single crystal X-ray crystallography studies may follow soon.

Finally, there is also the possibility that NU-1000 could be commensurately or incommensurately modulated.

## Appendix D: Comments on the experimental studies of Liu and coworkers [4,5]

Plane and frieze symmetry groups possess crystallographic origin conventions [21,31] that must be ensured to hold after alignment procedures of the raw data and symmetrized versions of the data whenever one wants to take full advantage of all of the mathematical relationships of 2D and 1D crystallography (see also appendix C1 in this context). When crystallographic origin conventions are ignored, as in ref. [4] for example, there is a good chance that subsequent symmetry classifications will be inconsistent. An example for such an inconsistency in the earlier version of ref. [4] is a supergroup ($p2mg$) possessing a smaller least-squares residual than one of its translationengleiche subgroups ($p211$). As equation (1) shows, this can never happen with good data that are processed correctly because $k_{more} > k_{less}$, both variables are positive integers, and $k_{less} > 1$.

Some of the translationengleiche subgroups of $p2mg$ and $p2mm$ possess for the walking human being least-squares residuals that seem to be too large to allow for the conclusion that either of these two groups can serve as the Kullback-Leibler best model on the basis of equation (1) even without having the benefit of a generalized noise level estimate. This could again be the result of ignoring the crystallographic origin convention for frieze symmetry groups.

One needs to comment finally on Liu's and coworkers' examples [4,5] that the time series of a gait pattern of any human being is not going to be perfectly periodic. Also the walking-person motif (that is periodic in time) possesses always site symmetry $1$ (i.e. identity after rotation by 360°) only and cannot feature any higher site symmetries. This is caused by the walking movement itself.

While standing still, on the other hand, the human body features an approximate mirror plane symmetry (in 3D). Unless a camera is not oriented perpendicular to the normal of this mirror plane, any photo or movie sequence of an upright standing person will not feature a mirror line symmetry (in 2D). While walking, the approximate mirror symmetry of a human being is "transformed" into a glide-line symmetry that possesses a translation component which is periodic in time but no site symmetries higher than $1$. (Good examples for this are the traces that a walking human being leaves in sand or freshly fallen snow.)

The frieze symmetry group of a walking person is, therefore, $p11g$, where there are no site symmetries other than the identity rotation. A time series of a walking human being can, however, only possess this frieze symmetry when the recording camera has been oriented so that its axis is perpendicular to the approximate mirror plane of the still standing person. In all other orientations, the camera will record a time series with freeze symmetry $p111$, i.e. the equivalent of pure translation periodicity only. (This is somewhat analogous to a walking person that carries a very heavy bag on her or his right shoulder. The traces of the right foot in freshly fallen snow will then be much deeper than the traces of the left foot and the $p11g$ symmetry is reduced to pure translation symmetry.)

All point symmetries of the walking human being motif higher than $1$ that are implied by frieze symmetry classifications higher than $p111$ and $p11g$ in the results of Liu and coworkers [4,5] are, therefore, actually genuine pseudo-symmetries [46].

Nevertheless, the studies of Liu and coworkers are valuable because the standard deviations of the mean values of the intensity of the pixels that collectively form a time-repeat unit decrease with the square root of the number of repeats when the noise is of the Gaussian type. A large reduction of the effects of the noise in a periodic time-repeat unit can, thus, be obtained when a very long time series is processed.

These kinds of studies should, however, not be considered as constituting genuine applied crystallography studies as there are only two possible outcomes when genuine pseudo-symmetries are not mistaken for genuine symmetries that form a crystallographic symmetry group.

Plane CSL grain boundaries in edge on-projections, as mentioned in the Introduction and Background section and appendix A, are, on the other hand, on bi-crystallography theory grounds [37] well described by frieze symmetry groups. Parts of the translation periodic motifs typically possess site symmetries higher than $1$ so that frieze symmetries higher than $p111$ and $p11g$ result.

27